\documentclass[twocolumn,english,aps,pra,floatfix,showpacs]{revtex4}
\usepackage{ae,aecompl}
\usepackage[T1]{fontenc}
\usepackage[latin9]{inputenc}
\usepackage{verbatim}
\usepackage{amsmath}
\usepackage{graphicx}
\usepackage{amssymb}

\makeatletter

\providecommand{\tabularnewline}{\\}

\@ifundefined{textcolor}{}
{%
 \definecolor{BLACK}{gray}{0}
 \definecolor{WHITE}{gray}{1}
 \definecolor{RED}{rgb}{1,0,0}
 \definecolor{GREEN}{rgb}{0,1,0}
 \definecolor{BLUE}{rgb}{0,0,1}
 \definecolor{CYAN}{cmyk}{1,0,0,0}
 \definecolor{MAGENTA}{cmyk}{0,1,0,0}
 \definecolor{YELLOW}{cmyk}{0,0,1,0}
 }

\usepackage{ae}
\usepackage{aecompl}

\makeatletter

\newcommand{\norm}[1]{\left\Vert #1 \right\Vert}
\newcommand{\abs}[1]{\left\vert#1\right\vert}

\newcommand{\ket}[1]{\left\vert#1\right\rangle}

\@ifundefined{definecolor}
 {\@ifundefined{definecolor}
 {\usepackage{color}}{}
}{}

\usepackage[
colorlinks=true,linkcolor=red,citecolor=green,urlcolor=blue,
pdfstartview=FitH,
bookmarksopen=true,bookmarksopenlevel=0,
plainpages=false,pdfpagelabels,
pagebackref=false,
pdftoolbar=false]{hyperref}

\makeatother

\usepackage{babel}

\begin{document}

\title{Dynamical Quantum Error Correction of Unitary Operations
with Bounded Controls}

\author{Kaveh Khodjasteh}
\author{Lorenza Viola}

\affiliation{{Department of Physics and Astronomy, Dartmouth
College, 6127 Wilder Laboratory, Hanover, New Hampshire 03755, USA}}

\begin{abstract}
Dynamically corrected gates were recently introduced {[}Khodjasteh and
Viola, Phys. Rev. Lett. \textbf{102}, 080501 (2009){]} as a tool to
achieve decoherence-protected quantum gates based on open-loop
Hamiltonian engineering. Here, we further expand the framework of
dynamical quantum error correction, with emphasis on elucidating under
what conditions decoherence suppression can be ensured while
performing a generic target quantum gate, using only available
bounded-strength control resources. Explicit constructions for
physically relevant error models are detailed, including arbitrary
linear decoherence and pure dephasing on qubits. The effectiveness of
dynamically corrected gates in an illustrative non-Markovian spin-bath
setting is investigated numerically, confirming the expected fidelity
performance in a wide parameter range. Robutness against a class of
systematic control errors is automatically incorporated in the
perturbative error regime.
\end{abstract}

\date{\today}


\maketitle

\section{Introduction}

The technological advances promised by quantum information processing
(QIP) science require an unprecedented level of control in
characterizing, manipulating, and measuring quantum systems
\cite{DiVincenzo-Overview}.  In reality, no quantum system may be
perfectly isolated from its surrounding environment and available
means for control and measurement are unavoidably limited. As a
results, deviations from the intended dynamical evolution arise, which
are collectively referred to as `errors'. Active quantum error
correction (QEC) methods aim to detect and counteract the effects of
errors during both quantum storage and processing of information via
non-trivial quantum gates \cite{NielsenBook}. Given limited control
resources, accurate quantum computation (QC) on an arbitrarily large
(scalable) QI processor is only possible if QEC is implemented
fault-tolerantly, that is, more errors is removed than is generated
when operations are themselves imperfect. The theory of fault-tolerant
QEC (FTQEC) guarantees that this is possible in principle, provided
that the \emph{error per gate} (EPG) is below a minimum threshold
\cite{Kitaev-Algorithms,PreskillRel,Knill-Resilient}. Unfortunately,
while architectures which can tolerate EPGs comparable with
experimentally achieved values have been proposed
\cite{Steane03,Knill-Noisy}, the corresponding resource overheads
remain prohibitive for existing QIP test-beds. Thus, QEC strategies
which can effectively reduce errors \emph{at the physical level} are
both crucial for boosting control fidelities in current quantum
devices and, in the long run, for enabling fault-tolerant QC.

While traditional QEC methods achieve information protection by
resorting to suitable redundant encodings and non-unitary quantum
operations, \emph{dynamical quantum error correction} (DQEC) aims to
counteract the effects of decoherence in a quantum system in a purely
`open-loop' fashion, that is, solely based on suitable unitary
(Hamiltonian) control.  Historically, dynamical procedures for
effectively removing unwanted couplings and/or evolution by
time-dependent `coherent averaging' have decades of tradition in
nuclear magnetic resonance (NMR) spectroscopy \cite{HaeberlenBook},
resulting in an unmatched degree of coherent control in liquid-state
nuclear-spin qubits \cite{Cory-Overview1,Ryan08}.  In particular,
dynamical decoupling (DD) sequences in NMR are early examples of DQEC
for a \emph{closed} quantum system, specifically designed to
`time-suspend' the underlying spin evolution by achieving a net
trivial (identity) propagator, or \textsc{noop} gate. In the QIP
context, DD methods have evolved into a powerful and versatile
framework for achieving decoherence suppression in \emph{non-Markovian
open} quantum systems in a variety of control settings
\cite{Viola1998,Viola1999Dec}.  While recent theoretical developments
include the exploitation of randomized design
\cite{Viola2005Random,Santos2006}, and the construction of
concatenated \cite{CDD} and optimal \cite{UDD} DD sequences, the
growing practical significance of DQEC methods is witnessed by the
intense parallel effort in the experimental QIP community. Following
simple proof-of-principle demonstrations in a single-photon
polarization interferometer \cite{Berglund2000}, DD in its simplest
(so-called `bang-bang') form has been successfully implemented by now
in solid-state nuclear quadrupole qubits \cite{Fraval}, fullerene
qubits \cite{Morton:2006}, coupled electron-nuclear systems
\cite{Morton:2008}, flying polarization qubits \cite{Vitali2008}, and,
most recently, trapped ions \cite{BiercukDD}.

Although the advances mentioned above demonstrate the usefulness of
DQEC toward removing unwanted decoherence \emph{in between} control
pulses and achieving a robust \textsc{noop} implementation,
full-fledged DQEC demands that decoherence be removed both \emph{in
between and during} control pulses, while implementing a robust
non-trivial (non-identity) quantum gate. \emph{Dynamically corrected
gates} (DCGs) were introduced in Ref. \onlinecite{khodjasteh-2008}
precisely to meet this goal.  As opposed to readily available
`primitive gates,' which effect a unitary operation on the system with
a `bare' EPG, each DCG consists of a suitably designated sequence of
control inputs in such a way that the desired target transformation is
effected with a net reduced EPG. DCGs are inspired by both
composite-pulse techniques for reducing classical control errors (in
particular, so-called `fully compensating' or `universal rotation'
sequences)
\cite{Levitt1986,Jones-Composite,Brown2004,KhanejaUR,Hill2007} as well
as strongly-modulating pulses
\cite{Fortunato-Control,Boulant03,KhanejaGRAPE} for high-fidelity
`soft' pulses in NMR, while differing from either strategy in
important ways. Specifically, DGCs compensate for error evolution due
to a \emph{dynamical quantum environment over which neither
quantitative knowledge nor direct control is assumed}. Although DCG
constructions are useful only in the short gating time/weak coupling
limit and cannot counter stochastic control errors, a key advantage is
that no extra qubits or measurement capabilities are needed, thanks to
the underlying open-loop control design. On the one hand, DCGs relate
to ongoing effort on improving pulse-shaping methods in order to
better accommodate physical constraints and/or increase robustness
against different errors. In particular, DD-inspired constructions for
pulses capable to `self-refocus' unwanted couplings have been recently
investigated in \cite{Pasini07,PryadkoQuiroz,Sengupta05}.  These
efforts nonetheless consider a specific control task (\textsc{noop}
implementation) and/or do not incorporate the effect of generic
quantum errors. DCGs, on the other hand, are applicable to any desired
unitary operation, and have the flexibility to handle a generic class
of non-Markovian error models on qubits. Furthermore, DCGs are
designed having high portability in mind, so that they can replace the
already available gates whenever appropriate (mild) compatibility
conditions are met.  Substituting primitive gates with DCGs can then
significantly lower physical EPGs, ultimately paving the way for more
practical FTQEC architectures.

In this paper, we expand our analysis of DCG constructions with the
twofold goal of describing in detail the mathematical groundwork
underlying the results of \cite{khodjasteh-2008}, as well as further
demonstrating the flexibility and potential of DCGs for high-fidelity
quantum control in open quantum systems. The content is organized as
follows. In Sec. \ref{sec:setup}, we formulate the DCG synthesis
problem in a control-theoretic setting, and describe the mathematical
assumptions and concepts that are relevant to the subsequent steps. In
Sec. \ref{sec:DDNOOP}, we review the fundamentals of DD methods based
on realistic bounded-strength controls, within so-called
\emph{Eulerian DD} \cite{Viola2003Euler}. In particular, we
re-interpret the latter as a special DCG aimed at quantum state
preservation, and prove a `No-Go' theorem that restricts the
portability of `black-box' DCGs beyond the \textsc{noop}
gate. Sec. \ref{sec:dcgbeyond} shows how to evade the No-Go theorem
and describes general DCG constructions in detail. Special attention
is paid to elucidating the compatibility requirements that emerge for
physically relevant qubit error models, including arbitrary linear
decoherence and pure dephasing. In Sec. \ref{sec:sim}, a concrete
spin-bath decoherence scenario is used as a case study to numerically
validate DCGs and to gain insight into open-system and control
features which affect their performance. In Sec. \ref{sub:extenapp},
we elaborate on the effects of additional error evolution induced by
control non-idealities and/or internal unitary drifts. A summary of
the main findings and open questions conclude in Sec. \ref{theend}.

\section{Control-theoretic setting for dynamical error correction}
\label{sec:setup}

Consider an isolated quantum system $S$ with an internal Hamiltonian
$H_{S}$ which can be controlled through the application of a
time-dependent Hamiltonian $H_{\text{ctrl}}(t)$. This control enables
us to perform certain quantum tasks on $S$. For example, by turning
off $H_{S}$ completely, an arbitrary state of $S$ may be
preserved. Complex control tasks such as unitary quantum gates and
algorithms composed of such gates may also be performed by modulating
$H_{S}$ through $H_{\text{ctrl}}(t)$.  In reality, however, $S$
interacts with an environment system (referred to as the `bath' $B$
henceforth) via an interaction Hamiltonian $H_{SB}$.  In the presence
of $H_{SB}$, a designated control sequence on $S$ is not sufficient to
accurately implement a desired evolution, due to non-unitary
decoherence errors that are induced once the bath degrees of freedom
are traced out.

Let, as usual, the initial joint state of $S$ and $B$ be separable,
with
$\rho^{\text{in}}=\rho_{S}^{\text{in}}\otimes\rho_{B}^{\text{in}}$,
and
$\rho_{S}^{\text{in}}=|\psi_{\text{in}}\rangle\langle\psi_{\text{in}}|$.
Ideally, the action of a target unitary gate $Q$ effected over a
gating interval $[0,T]$ should result in a final state of the form
$\rho_{SB}^{\text{fin}}(T)=Q^{\dagger}
\rho_{S}^{\text{in}}Q\otimes\rho_{B}^{\text{fin}}(T)$, for some
(irrelevant) final state $\rho_{B}^{\text{fin}}(T)$ of the bath. In
contrast, the actual action of the gate in the presence of $H_{SB}$
results in the combined joint state $\rho_{SB}^{\text{err}}(T)$, and
the system ends up in an erroneous reduced state
$\rho_{S}^{e}(T)=\text{Tr}_{B}(\rho_{S}^{\text{err}}(T))\ne
Q^{\dagger}\rho_{S}^{\text{in}}Q$.  While different measures may be
envisioned to quantify the deviation between the intended and the
actual evolution, the resulting EPG will be proportional to the gating
time $T$, as well as to the overall `error strength,' appropriately
defined. In essence, DCG constructions aim to \emph{improve the
scaling of physical EPGs with the gating time in the limit where the
latter is sufficiently small}, so that a perturbative approach is
meaningful. That is, a gate $Q_{\text{DCG}}$ is an $\ell$-\emph{th}
\emph{order DCG} if it obeys
\begin{equation}
\frac{\text{EPG}^{[\ell]}}{\text{EPG}^{[0]}}=O(T^{\ell}),\;\;\;\ell
\geq1,\label{dcgl}
\end{equation}
\noindent
where the asymptotic $O$-notation is used to signify the limit where
the zero-th order bare EPG is close to 0. If desired, the above
defining condition for a DCG may be expressed in terms of gate
fidelity improvement once a specific fidelity metric is chosen and is
related to the EPG \emph{amplitude} in Eq. (\ref{dcgl}). Let, for
instance, $f_{Q}$ denote the worst-case fidelity in implementing $Q$
\cite{NielsenBook},
\begin{equation}
f_{Q}=\min_{|\psi_{\text{in}}\rangle}\,\text{Tr}
\sqrt{\sqrt{\rho_{S}^{\text{err}}}\,{\rho_{S}^{\text{fin}}}\,
\sqrt{\rho_{S}^{\text{err}}}}.\label{minfid}
\end{equation}
\noindent
By taking the `infidelity' $(1-f_{Q})$ as a measure of the gate error
\emph{probability}, a $\ell$-th order DCG implies an improvement ratio
on the order of \cite{Brown2004,Terhal-FT,Viola2005Random}
\begin{equation}
r^{[\ell]}=\frac{1-f_{Q}\;\:}{1-f_{Q_{\text{DCG}}}^{[\ell]}}
=O(T^{-2\ell}),\label{ratio}
\end{equation}
\noindent
where $f_{Q_{\text{DCG}}}$ is the fidelity of implementation of a
sequence of primitive gates whose fidelities are all similar and given
by $f_{Q}$.

\par In this paper we will only focus on first-order DCGs ($\ell=1$),
by postponing the construction and characterization of higher-order
DCGs to a separate forthcoming analysis \cite{khodjasteh-2009}.

\subsection{Error model assumptions}
\label{sub:errassu}

Let $S$ consist of $n$ qubits, and let $B$ represent another quantum
system over which \emph{knowledge is limited and no direct control is
available}. The free evolution of $S$ and $B$ in the combined Hilbert
space $\mathcal{H}=\mathcal{H}_{S}\otimes\mathcal{H}_{B}$ is described
by a bare internal Hamiltonian $H_{\text{int}}$ of the form
\begin{equation} H_{\text{int}}=H_{S}\otimes I_{B}+H_{SB}+I_{S}\otimes
H_{B},\label{eq:Hall}
\end{equation}
\noindent
where $I_{S}$ ($I_{B}$) is the identity operator in the operator space
${\cal B}(\mathcal{H}_{S})$ {[}${\cal B}(\mathcal{H}_{B}$){]},
respectively. Otherwise obvious tensor products components will be
dropped from now on; \emph{e.g.}, $H_{B}$ will be understood as
$I_{S}\otimes H_{B}$.  The interaction Hamiltonian $H_{SB}$ is
responsible for decoherence effects whereby initially separable states
of $S$ and $B$ become entangled and loose their capacity for QIP. The
typical time scale over which such effects become significant (loosely
referred to as the `decoherence time') depends in general on both
$H_{SB}$ and various features of the bath, including
$\rho_{B}^{\text{in}}$ \cite{RevModPhys.75.715,Breuer}.  Without loss
of generality, $H_{SB}$ can be expanded as
$H_{SB}=\sum_{\gamma}S_{\gamma}\otimes B_{\gamma}$, where
$\{S_{\gamma}\}$ is an (Hermitian) operator basis for ${\cal
B}(\mathcal{H}_{S})$, such as arbitrary products of Pauli operators on
the $i$-th qubit, denoted by $\{S_{\alpha}^{(i)}\}$
($\alpha\in\{I,X,Y,Z\}\equiv\{0,1,2,3\}$), and $B_{\gamma}$ being
non-zero but otherwise unspecified bath operators. In most
circumstances, the dominant terms in $H_{SB}$ belong to a restricted
subspace of all possible coupling operators. In particular, the
\emph{linear decoherence model} is defined by letting:
\begin{equation}
H_{SB}^{\text{lin}}=\sum_{i=1}^{n}\sum_{\alpha=1}^{3}S_{\alpha}^{(i)}\otimes
B_{\alpha}^{(i)}.
\label{eq:Hsbsingle}
\end{equation}
\noindent
Physically, the class of error models described by
Eq. (\ref{eq:Hsbsingle}) allows for arbitrary combinations of phase-
and amplitude- damping processes involving a \emph{generic} axis and
degree of correlations between different qubits (from independent to
fully collective linear interactions) \cite{Knill-NS}. In DQEC
approaches, it is useful to associate an \emph{error subspace}
$\Omega_{e}\subset{\cal B}({\cal H})$ to the set of error generators
we wish to suppress. For linear decoherence,
\begin{equation}
\Omega_{e}=\text{span}\,\{S_{\alpha}^{(i)}\otimes B\,|
\,\forall\alpha,i;\; B\in{\cal B}({\cal H}_{B})\}
\equiv\Omega_{e}^{\{1\}}\label{eq:Hsbsingle1}
\end{equation}
\noindent
includes arbitrary system-bath operators that induce single-qubit
errors on $S$. Also notice that the strength of the system-bath
interaction Hamiltonian is constrained to be sufficiently weak in
order for both QEC and DQEC to be useful in practice. This motivates
assuming that both ${H_{SB}}$ and ${H_{B}}$ are \emph{bounded} in an
appropriate norm, although otherwise potentially unknown. The
unitarily invariant operator norm $\norm{\cdot}$ \cite{bhatiaBook}
will be used throughout our analysis.

If $H_{SB}$ is not present, a desired unitary operation $Q$ on $S$ can
be generated, as mentioned earlier, by adjoining a semi-classical
controller, described by a time-dependent Hamiltonian
$H_{\text{ctrl}}(t)$.  Depending on the system and control
specifications, certain components of the internal system Hamiltonian
might be essential to ensure universal controllability, whereas other
will induce unintended (unitary) error evolution. In general, we may
thus decompose $H_{S}=H_{S,g}+H_{S,e}$, where the separation into the
gating and error contribution depends on the target transformation
$Q$; for instance, $H_{S}\equiv H_{S,e}$ for \textsc{noop},
$Q=I_{S}$. In the presence of $H_{SB}$, let the \emph{total error
Hamiltonian} be defined by
\begin{equation}
H_{e}=H_{S,e}+H_{SB}+H_{B},
\label{errHam}
\end{equation}
\noindent
and, correspondingly, we represent the control action in terms of a
gating Hamiltonian of the form
\begin{equation}
H_{\text{gate}}(t)\equiv H_{\text{gate}}(t)\otimes
I_{B}=H_{\text{ctrl}}(t)+H_{S,g}.
\label{naive}
\end{equation}
\noindent
In principle, a perfect implementation of $Q$ could still be ensured
through infinitely strong, instantaneous gating Hamiltonians, during
which $H_{e}$ has no time to act. However, this ideal scenario cannot
be achieved in reality, although it may be realized approximately.
Even if $H_{SB}\ne0$, it may be useful to have a way for comparing
operators in ${\cal B}(\mathcal{H})$ which differ only through
directly uncontrollable pure-bath terms such as $H_{B}$. This can be
accommodated by defining equivalence classes of operators: We say that
\emph{$X$ is equal to $Y$ modulo the bath} (mod$B$ for short) iff
their difference is a \emph{pure-bath operator}, \[ X=Y\text{
mod$B$}\Longleftrightarrow\ \exists X_{B}:\ X-Y=I_{S}\otimes X_{B},\]
where $X_{B}$ is an operator acting on $B$. For example,
$H_{gate}(t)+H_{B}=H_{gate}(t)\text{ mod$B$}$.  With this in mind, we
use the words \emph{ideal} or \emph{desired} to refer to cases with
$H_{e}=0$ mod$B$, where gates can be perfectly implemented by a
suitable closed-system protocol $H_{\text{gate}}(t)$ over time
$T$. Furthermore, we will primarily focus on open-system
\emph{decoherence errors}, that is, we will treat $H_{SB}$ as the
leading source of errors in Eq. (\ref{errHam}), by assuming that:

\texttt{(a1)} \texttt{Perfect control assumption}: No additional
errors are introduced by the controller;

\texttt{(a2)} \texttt{Driftless system assumption}: No additional
errors are introduced by the system's internal evolution.

Assumption \texttt{(a1)} implies that the applied control Hamiltonian
$H_{\text{ctrl}}(t)$ is perfect, subject to operational constraints to
be specified later (Sec. \ref{sub:contobj}). While realistic controls
will suffer in general of both systematic and stochastic (random)
imperfections, these errors are intrinsically \emph{classical} in
nature. Likewise, assumption \texttt{(a2)} is partly justified by the
fact that $H_{S}$ is fully known and, in the closed-system limit,
complete control over $S$ is available. Once DCG constructions are
found under the above simplifying assumptions, it is at least in
principle conceivable that simultaneous compensation of classical and
quantum (unitary and decoherence) errors could be achievable by
suitably merging DCG constructions with existing robust control
techniques, although several non-trivial aspects may remain. While
additional discussion is deferred to Sec. \ref{sub:extenapp}), the
impact of control imperfections will be also numerically assessed in
Sec. \ref{sec:sim}.

\subsection{Control assumptions and error measure}
\label{sub:contobj}

Our goal is to use available control resources to generate unitary
gates on $S$ while minimizing the sensitivity of the gate action to
the error Hamiltonian $H_{e}$. In practice, allowed manipulations will
be restricted by a number of constraints, which can be identified by
writing
\begin{equation}
H_{\text{ctrl}}(t)=\sum_{u}h_{u}(t)V_{u}\equiv\sum_{u,\gamma}
f_{\gamma}^{u}(t)S_{\gamma},
\label{control}
\end{equation}
\noindent
in terms of a subset of \emph{admissible control Hamiltonians}
$\{V_{u}\}$ and of \emph{control inputs} $\{h_{u}(t)\}$, the latter
being real-valued controllable functions of time. Limited
`pulse-shaping' capabilities may additionally restrict the values
attainable by the control inputs $h_{u}(t)$ and/or their
derivatives. Two main constraints will be incorporated into the
present DCG design:

\texttt{(c1)} \texttt{Finite power}, which requires bounded control
amplitudes, $h_{u}(t)\leq h_{\text{max}}$;

\texttt{(c2)} \texttt{Finite bandwidth}, which requires a bounded
Fourier spectrum, hence a minimum time scale $\tau_{\text{min}}$ for
modulation.

\par A `primitive gate' will refer to a physically available gate used
to generate more complex transformations. In the simplest case, a
primitive gate may be implemented by turning on and off a single
control Hamiltonian in the available set, subject to the above
assumptions.  A `control block' will likewise refer to a time interval
during which multiple primitive gates are applied sequentially in a
predetermined manner. From the standpoint of universal QC, two
different scenarios arise depending on whether the set of switchable
control Hamiltonians $\{V_{u}\}$ in Eq. (\ref{control}) supports a
\emph{universal set of single- and two- qubit gates}. If this is the
case, $\{V_{u}\}$ generates a dense subset in SU$(2^{n})$, thus
arbitrary unitary gates on $S$ may (ideally) be approximated to a
desired accuracy by turning on and off primitive Hamiltonians in the
set.  Note that if no contribution from the internal Hamiltonian $H_S$
is needed to achieve universality ($H_{S,g}=0$), then to the purpose
of DCG design we may effectively treat $H_S$ as a pure error
($H_S\equiv H_{S,e}$) and the problem as effectively driftless, in
line with \texttt{(a2)}.

\par Let $Q$ denote a quantum gate generated through the application
of $H_{\text{ctrl}}(t)$ over the time interval $[0,T]$. Recalling Eq.
(\ref{naive}), the gating propagator $U_{\text{gate}}(t,0)$,
$t\in[0,T]$, can be defined in general as
\begin{eqnarray*}
U_{\text{gate}}(t,0) & = &
T_{+}\Big[\exp\Big(-i\int_{0}^{t}H_{\text{gate}}(s)ds\Big)\Big],
\end{eqnarray*}
\noindent
such that $U_{\text{gate}}(T,0)=Q$.  We use $U_\text{gate}(t)$ to
denote $U_\text{gate}(t,0)$ when there is no ambiguity about the
starting time of the gate.  The overall propagator for the system plus
bath is given by \[ U(T,0)=T_{+}\Big[\exp\Big(-i\int_{0}^{T}
(H_{\text{gate}}(s)+H_{e})ds\Big)\Big].\] A natural way to isolate the
effect of the error Hamiltonian is to effect a transformation to a
frame that `toggles' with the applied control, that is, to move to the
interaction picture generated by $U_{\text{gate}}(t,0)$:
\begin{eqnarray} U(t,0) & = &
U_{\text{gate}}(t,0)e^{-i\Phi(t,0)},\\ e^{-i\Phi(t,0)} & \equiv &
T_{+}\Big[\exp\Big(-i\int_{0}^{t}H_{e}(s,0)ds\Big)\Big],
\label{eq:errortoggle}
\end{eqnarray}
\noindent
where the `modulated' error Hamiltonian reads
\begin{eqnarray}
H_{e}(t,0) & = &
U_{\text{gate}}(t,0)^{\dagger}H_{e}U_{\text{gate}}(t,0).
\label{eq:Het}
\end{eqnarray}
\noindent
The Hermitian `error action operator' $\Phi(T,0)$ defined in Eq.
(\ref{eq:errortoggle}) is directly related to the effective
Hamiltonian which describes the joint evolution in the toggling
frame. Mathematically, $\Phi(T,0)$ provides, up to pure bath terms, a
measure of the overall deviation of the actual evolution of $S$ from
the desired unitary evolution $Q$. Since the norm of $\Phi(T,0)$ may
be used to lower-bound the minimum achievable fidelity
\cite{Lidar-Bounds}, the latter may be used as a natural EPG metric
for DCG constructions {[}Eq. (\ref{dcgl}){]}, and the associated
quantum-control problem may be viewed as the minimization of
$\Phi(T,0)$, up to pure-bath terms. The solution obtained in
\cite{khodjasteh-2008} and further investigated here is perturbative
and analytic: We guarantee that $\Phi(T,0)\sim\text{EPG}^{[1]}$ is of
\emph{second order} in $\tau_{\text{min}}$, and refer to this
perturbative cancellation as `correcting first-order errors,' while
similarly referring to the residual (asymptotically smaller) terms in
$\Phi$ as `uncorrected'.

\subsection{Tools for error analysis}
\label{sec:tools}

Our next step is to describe how the error action $\Phi$ can be
evaluated and approximated for complex evolutions involving a
composite control block. Suppose that the interval $[0,T]$ is
decomposed into two segments $[0,T_{1}]$ and $[T_{1},T]$. By using the
composition property of unitary propagators and, for each interval,
the separation between intended and error contribution, we have
\begin{eqnarray}
e^{-i\Phi(T,0)} & = &
U_{\text{gate}}^{\dagger}(T,0)U_{\text{gate}}(T,T_{1})
e^{-i\Phi(T,T_{1})}U(T_{1},0)\nonumber \\ & = & \exp(-i\,
U_{\text{gate},1}^{\dagger}\Phi_{2}U_{\text{gate},1})\exp(-i\Phi_{1}),
\label{eq:ercombine}
\end{eqnarray}
\noindent
where
\begin{eqnarray*} & &
U_{\text{gate},1}=U_{\text{gate}}(T_{1},0),\\ & &
e^{-i\Phi_{1}}=e^{-i\Phi(T_{1},0)},\;\;\;
e^{-i\Phi_{2}}=e^{-i\Phi(T,T_{1})}.
\end{eqnarray*}
\noindent
Eq. (\ref{eq:ercombine}) allows to recursively combine segments of
evolution entering a composite unitary gate and to calculate the EPG
associated with the combined sequence in terms of EPG of each
segment. Physically, this is equivalent to transforming to consecutive
toggling frames. In this sense, the error associated with each
individual segment can be isolated and computed, provided that the
control propagator during the segment and the control path applied up
to the beginning of the segment are known.

Generally, when $N$ gates $Q_{1},\cdots,$ $Q_{N}$ are applied in
succession, the error operator associated with the combination
$A=Q_{N}Q_{N-1}\cdots Q_{1}$ is given by $\Phi_{A}$, with
\begin{eqnarray}
\exp(-i\Phi_{A}) & = &
e^{-iP_{N-1}^{\dagger}\Phi_{Q_{N}}P_{N-1}}\cdots
\label{eq:combinegates}\\
& &
e^{-iP_{1}^{\dagger}\Phi_{Q_{2}}P_{1}}e^{-iP_{0}^{\dagger}
\Phi_{Q_{1}}P_{0}},\nonumber
\end{eqnarray}
\noindent
where the `partial' gating propagators up to segment $i$ along the
path are given by \[ P_{i}=Q_{i}Q_{i-1}\cdots Q_{1},\;
i=1,\ldots,N-1;\;\; P_{0}=I_{S}.\] Notice that from
Eq. (\ref{eq:combinegates}), in the limit of infinitesimal segments,
we recover the toggling frame expression for $\Phi(T,0)$, as given in
Eqs. (\ref{eq:errortoggle})-(\ref{eq:Het}). Eq. (\ref{eq:combinegates})
also lends itself naturally to approximation. If the errors associated
with the individual gates $Q_{i}$ are known, then the (discrete-time)
Magnus expansion provides a systematic (albeit computationally
demanding for higher orders) series expansion for $\Phi_{A}$:
\begin{eqnarray}
\Phi_{A} & = &
\sum_{i=1}^{\infty}\Phi_{A}^{[i]},\nonumber \\ \Phi_{A}^{[1]} & = &
\sum_{i=1}^{N}P_{i-1}^{\dagger}\Phi_{Q_{i}}P_{i-1},\;\;
\label{eq:PhiA1}\\
\big\Vert\sum_{i=2}^{\infty}\Phi_{A}^{[i]}\big\Vert & = &
O(\max_{i}\norm{\Phi_{Q_{i}}}^{2}),\nonumber
\end{eqnarray}
\noindent
and the above estimate for higher-order corrections
\cite{Khodjasteh-Hybrid} holds as long as (absolute) convergence is
ensured, that is, if $N\max\norm{\Phi_{Q_{i}}}<\pi$.  The
(continuous-time) Magnus expansion can similarly be applied to
approximate errors directly in the toggling frame
\cite{Viola1999Dec,Viola2003Euler} or to estimate the individual EPGs
$\Phi_{Q_{i}}$. Starting with Eq. (\ref{eq:errortoggle}), in analogy
to Eqs. (\ref{eq:PhiA1}) we have: \begin{eqnarray} \Phi(t,0) & = &
\sum_{i=1}^{\infty}\Phi^{[i]}(t,0),\nonumber \\ \Phi^{[1]}(t,0) & = &
\int_{0}^{t}H_{e}(s,0)ds,\label{eq:PhiA2}\\
\big\Vert\sum_{i=2}^{\infty}\Phi^{[i]}(t,0)\big\Vert & = &
O(\norm{tH_{e}}^{2}),\nonumber \end{eqnarray} as long as
$t\norm{H_{e}}<\pi$. In what follows, we will use $X^{[n]}$ to denote
the $n$-th Magnus expansion term for the operator $X$ and $X^{[n+]}$
to denote the sum of $n$-th and higher-order terms.  Furthermore, when
depicting the control inputs for a gate, we will assume that the
gating period starts at the time $t_{0}=0$, unless otherwise
stated. This is possible due to the fact that when individual gates
are considered, the EPG does not depend on the initial gating time
unless the physical error Hamiltonian in Eq. (\ref{errHam}) is itself
explicitly time-dependent.

\section{Dynamically corrected NOOP}
\label{sec:DDNOOP}

The `no operation' (\textsc{noop}) gate plays an especially important
role from both a quantum-control and a QIP standpoint. On the one
hand, even in a closed-system setting, the ability to un-do the effect
of the drift and realize an arbitrary closed control trajectory is a
prerequisite for complete controllability \cite{D'Alessandro}.  On the
other hand, preserving an arbitrary quantum state on all or part of a
quantum circuit for a desired storage time is an essential subroutine
for complex quantum algorithms. As such, achieving the \textsc{noop}
gate is an important benchmark for DD methods. In turn, DD provides us
with an inner combinatorial structure for DCGs which goes beyond the
\textsc{noop} gate itself.

\subsection{Basics of dynamical decoupling}

DD is the most prevalent DQEC strategy. In essence, DD schemes subject
the system to either sequences of infinitely strong instantaneous
`$\delta$-pulses' (bang-bang DD, BB DD) or bounded-strength control
modulations (Eulerian DD, EDD), in such a way that undesired evolution
induced by the bath is {\em coherently averaged out} over a time scale
longer than a minimum control time scale $T_{c}$. While DD performance
can depend heavily upon the details of the sequence used, a basic
requirement for averaging is that $T_{c}$ be sufficiently short as
compared to the time scales of the erroneous dynamics to be
removed. While, as mentioned, a myriad of progressively more
elaborated DD schemes have emerged
\cite{Khodjasteh2004,Viola2005Random,Uhrig2007,Uys09}, we briefly
review here the original group-theoretic construction
\cite{Viola1999Dec} in a form suitable to our current scope.

\par Consider a finite group $\mathcal{G}=\{g_{i}\}_{i=1}^{d}$,
$d\equiv|{\cal G}|$, which acts through a faithful unitary
(projective) representation $\{G_{i}\}_{i=1}^{d}$ on $\mathcal{H}_{S}$
and through a corresponding adjoint representation on ${\cal B}({\cal
H}_{S})$. These representations can be extended trivially to the
combined Hilbert space $\mathcal{H}$ by using $\{G_{i}\otimes I_{B}\}$
as the representation of $\mathcal{G}$.  $\mathcal{G}$ is a
\emph{decoupling group} for a subspace $\Omega\subset{\cal B}({\cal
H}_{S})$ of (traceless) operators on $\mathcal{H_{S}}$ iff \[ \forall
E\in\Omega:\
\Pi_{\mathcal{G}}(E)=\sum_{i=1}^{d}G_{i}^{\dagger}EG_{i}=0,\]
\noindent
where $\frac{1}{d}\Pi_{\mathcal{G}}(\cdot)$ is the projection
superoperator onto the ${\cal G}$-invariant sector, and we refer to
each $G_{i}$ as a decoupling operator for $\Omega$. For example,
consider ${\cal G}=\mathbb{Z}_{2}$, represented by $\{I,X\}$ on the
state space ${\cal H}_{S}={\mathbb{C}}^{2}$ of one qubit. This group
is a decoupling group for the subspace $\Omega_{\perp
X}=\text{span}\,\{Z,Y\}$, thus it can correct error Hamiltonians in
the \emph{correctable error space} $\Omega_{c}=\{Z\otimes
B_{Z}+Y\otimes B_{Y}+I_{S}\otimes B_{0}\}$, regardless of the
arbitrary tensor components $B_{\alpha}$. In general, if $\Omega_{e}$
is the subspace of system-bath operators generated by all errors we
wish to remove, a good DD group ensures that
$\Omega_{c}\supseteq\Omega_{e}$.

\par Consider next a control propagator $U_{\text{DD}}(t)$ applied on
$d$ consecutive time intervals $I_{i}=[t_{i-1},t_{i}]$, where
$i=1,\ldots, d$, $t_{0}=0$, and $(t_{i}-t_{i-1})=\tau$, such that each
decoupling operator is implemented once, beginning from the identity
$I_{S}\equiv G_1$. That is,
\begin{equation}
U_{\text{DD}}(t)=\left\{ \begin{array}{ll} G_{i} & \text{if
}t\in[t_{i-1},t_{i}),\;\; i=1,\ldots,d,\\ I_{S} & \text{if
}t\notin(0,t_{d}).\end{array}\right.\label{bbdd}
\end{equation}
In principle, the discontinuous (in time) propagator
$U_{\text{DD}}(t)$ can be realized by a singular control Hamiltonian
that is switched on at $t_{i}$ with infinite strength, and
instantaneously rotates $S$ according to
$D_{i}=G_{i+1}G_{i}^{\dagger}$ for $i=1, \ldots, d-1$, and
$D_{d}=G_d^\dagger$. Notice that the propagator after $t_{d}$ is the
identity operator, which amounts to a \textsc{noop} gate associated
with the whole evolution. $S$ and $B$ undergo free evolution governed
by $H_{e}$ during $(t_{i-1},t_{i})$ with an error given by
$\Phi_{\tau}$, but $S$ is kicked at $t_{i}$ by $D_{i}$. Let
$T=d\,\tau$. The error associated with the combined sequence in
Eq. (\ref{bbdd}) can be obtained by using Eq. (\ref{eq:errortoggle}):
\begin{eqnarray}
\Phi_{\text{DD}} & = &
G_{d}\Phi_{\tau}G_{d}^{\dagger}\cdots
G_{2}\Phi_{\tau}G_{2}^{\dagger}G_{1}\Phi_{\tau}G_{1}\nonumber \\ & = &
\prod_{i=1}^{d}\exp(-i\tau
G_{i}^{\dagger}H_{e}G_{i})=\Phi_{DD}^{[1]}+\Phi_{DD}^{[2+]},\nonumber
\\ \Phi_{\text{DD}}^{[1]} & = &
\exp\Big(\!\!-i\tau\sum_{i=1}^{d}G_{i}^{\dagger}H_{e}G_{i}\Big)
=e^{-i\, \frac{T}{d}\Pi_{{\cal G}}(H_{e})},
\label{eq:phidd1}
\end{eqnarray}
\noindent
where the Magnus expansion of Eq. (\ref{eq:PhiA1}) has been used to
approximate $\Phi_{DD}$ in the leading powers of $\tau H_{e}$,
assuming that $T\norm{H_{e}}<\pi$. As long as $H_{e}\in\Omega_{e}$,
then $\Phi_{\text{DD}}^{[1]}=H_{B}=0\text{ mod}B$, thus the protocol
will produce a (first-order) error-corrected \textsc{noop} on $S$.
Notice that we could alternatively envision the DD sequence as a
sequence of (ideal) gates $D_{i}$ with errors $\Phi_{D_{i}}=0$ (mod
$B$) and primitive \textsc{noop} gates $F$ (consisting of free
evolution) with error $\Phi_{F}=\tau H_{e}$, applied as $D_{d}F\cdots
D_{2}FD_{1}F$, with a combined error given by
Eq. (\ref{eq:combinegates}). The uncorrected error associated with the
DD sequence is given by the higher-order Magnus terms
\cite{Viola1999Dec,Khodjasteh2005}.

\par Theoretically, the efficiency of the basic `periodic' BB DD
scheme based on (\ref{bbdd}) is limited by the requirements that (i)
the control time scale $T_{c}=d\tau$ be short enough for the Magnus
expansion to be applicable; and (ii) the higher-order error terms
$\Phi_{\text{DD}}^{[2+]}$ be sufficiently small. The impact of
higher-order corrections is especially important if repeated control
cycles are implemented, in order to achieve a net \textsc{noop} over a
desired (arbitrarily long) storage time. More sophisticated DD
techniques address these issues by invoking a number of design
improvements -- including recursive constructions
\cite{Khodjasteh2004}, randomization of the control path
\cite{Viola2005Random,Santos2008}, and optimization of the control
intervals \cite{Uhrig2007,Uys09}.  Presently, however, for the generic
linear decoherence model of Eq.  (\ref{eq:Hsbsingle}), no provably
optimal DD sequence exists; that is, unlike in the pure dephasing case
where decoherence occurs in a preferred basis, no {\em efficient}
high-order DD schemes are known. Furthermore, in complex systems of
interacting qubits, DD may be used to remove unwanted inter-qubit
couplings or to selectively enforce a desired dynamics. In such
scenarios, identifying a {\em minimal} DD group and its corresponding
physical representation is, likewise, a largely open problem in
combinatorics \cite{Wocjan2006}.

\subsection{Eulerian decoupling design}

A major simplification afforded by the BB setting is the separation
between the evolution due to the controller and the one due to the
internal open-system Hamiltonian $H_{\text{int}}$. Realistic
operations, however, always entail a bounded control amplitude,
thereby a finite duration. Even if devices exist where $\delta$-pulses
are an adequate first approximation, from a control-theory standpoint
it is highly desirable to recover the BB setting as a limiting case of
a formulation based on (non-singular) admissible controls. This is the
central motivation in EDD \cite{Viola2003Euler}.

\begin{figure}[tb]
\begin{centering}
\includegraphics[width=1.8in]{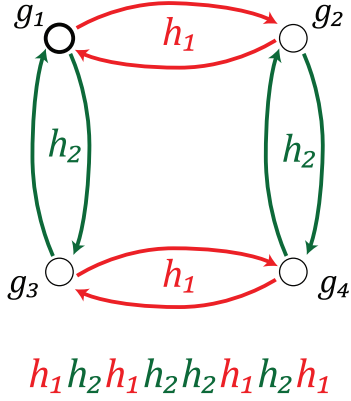} \par\end{centering}
\caption{(color online) The Cayley graph $G(\mathcal{G},\Gamma)$ with
$\mathcal{G}=\mathbb{Z}_2\otimes\mathbb{Z}_2=\{g_i\}_{i=1}^4=
\{(0,0),(0,1),(1,0),(1,1)\}$ and
$\Gamma=\{h_1,h_2\}=\{g_2,g_3\}=\{(0,1),(1,0)\}$ with an Eulerian path
of length $L=8$ depicted.  This group with the Cayley graph and the
Eulerian cycle depicted here will be used in Sec. \ref{sec:Examples}
to implement DCG constructions for the linear decoherence model.
\label{fig:cayley_abs}}
\end{figure}

Consider a set of generators $\Gamma=\{h_{j}\}_{j=1}^{m}$ for
$\mathcal{G}$, such that each element $g\in\mathcal{G}$ can be written
as an ordered product of $h_{i}$'s. The Cayley graph
$G(\mathcal{G},\Gamma)$ of $\mathcal{G}$ with respect to $\Gamma$
pictorially represents the generation of all elements of $\mathcal{G}$
through application of $h_{j}$: Each vertex of $G(\mathcal{G},\Gamma)$
represents a group element, and a vertex $g_{\ell}$ is connected to
another vertex $g_{\ell'}$ by a directed edge `colored' (labeled) with
the generator $h$ iff $g_{\ell'}=hg_{\ell}.$ The number of edges of
$G(\mathcal{G},\Gamma)$ is thus given by $L=dm$. It can be shown that
every Cayley graph possesses an \emph{Eulerian cycle} that visits
every edge exactly once.  Let ${\cal C}$ be an Eulerian cycle on
$G(\mathcal{G},\Gamma)$ that starts (and ends) at the identity (see
Fig. \ref{fig:cayley_abs} for a pictorial view).

This cycle can be associated to the control propagator
$U_{\text{EDD}}(t)$ which describes evolution in $L$ consecutive
intervals $I_{j}=[t_{j-1},t_{j}]$, that is,
\begin{equation}
U_{\text{EDD}}(t_{j})=F_{j}U_{\text{EDD}}(t_{j-1}),\;\;\;
j=1,\ldots,L,
\label{edd}
\end{equation}
\noindent
where $F_{j}$ is the group representation of the generator labeling
the $j$-th edge in ${\cal C}$. According to Eq. (\ref{edd}), the
control propagator follows the Eulerian path faithfully: The gating
Hamiltonian $H_{\text{gate}}(t)$ varies in such a way that \emph{each}
evolution segment labeled by a generator $h_{j}$ along the path is
implemented by a corresponding gate $F_{j}$. The evolution over
$T_{c}=L\tau$ thus consists of each generator $F_{j}\in\Gamma$ being
implemented precisely $d$ times. Suppose that the error associated
with each gate $F_{j}$ is given by $\Phi_{F_{j}}$, determined by the
manner in which $F_{j}$ is physically implemented: \[
\exp(-i\Phi_{F_{j}})=T_{+}\Big[\exp\Big(\!-i\int_{0}^{\tau}\!
U_{\text{gate},j}(t)^{\dagger}H_{e}U_{\text{gate},j}(t)\Big)\Big],\]
where we have assumed that $F_{j}$ is generated through the
application of a gating Hamiltonian $H_{\text{gate},F_{j}}(t)$ in the
interval $[0,\tau]$ (which can be shifted forward in time), with \[
U_{\text{gate},j}(t)=T_{+}\Big[\exp\Big(\!-i\int_{0}^{t}\!
H_{\text{gate},F_{j}}(s)ds\Big)\Big].\] The combined error associated
with the EDD sequence can be approximated by a Magnus expansion
{[}Eq. (\ref{eq:PhiA1}){]}:
\begin{eqnarray}
\Phi_{\text{EDD}}= \sum_{i=1}^{m}\Pi_{\mathcal{G}}
(\Phi_{F_{i}})+\Phi_{\text{EDD}}^{[2+]},
\label{eq:eddsum}
\end{eqnarray}
\noindent
with
$\norm{\Phi_{\text{EDD}}^{[2+]}}=O(\max_{j}\norm{\Phi_{F_{j}}}^{2})$.
In EDD, we require that the errors $\Phi_{F_{j}}$ associated with the
control generators belong to the error set $\Omega_{e}$ corrected by
$\mathcal{G}$. As long as this is the case, then
$\Phi_{\text{EDD}}^{[1]}=0$ mod$B$, and the EDD sequence performs a
\textsc{noop} with an asymptotically small error
$\Phi_{\text{EDD}}^{[2+]}$ compared to the uncorrected evolution.

Given arbitrary linear decoherence, we will show in
Sec. \ref{sec:Examples} how readily available control options for
realizing the generators of ${\cal G}$ can produce errors that indeed
belong to $\Omega_{e}$.  For a generic open system, it need not be
apparent which type of Hamiltonians and control inputs (if any) might
be used to generate EDD generators such that their respective errors
$\Phi_{F_{j}}$ are correctable by (some) ${\cal G}$, a question we
refer to as the `gate permissibility problem'. To summarize, provided
a set of permissible generator gates can be identified, EDD yields a
constructive solution to our main problem as defined in
Sec. \ref{sub:contobj}, with the desired target gate being the
\textsc{noop} gate.

\subsection{No-go theorem for black-box constructions}
\label{subsec:contobv}

An important feature of EDD is that the net error given by
Eq. (\ref{eq:eddsum}) is canceled \emph{irrespective} of how each gate
$F_{j}$ representing a decoupling generator is implemented in terms of
the control inputs, provided that (i) the same implementation is used
each time $h_{j}$ appears along the path and (ii) each error
$\Phi_{F_{j}}$ is correctable by ${\cal G}$. To state it differently,
\emph{no a priori relationship between the errors $\{\Phi_{F_{j}}\}$
is assumed}, as they cancel independently up to the first order
{[}Eq. (\ref{eq:eddsum}){]}.  In this sense, error correction in EDD
is {oblivious to the details of the control}. Unfortunately, as we
shall prove next, such `control-oblivious decoupling' can only
dynamically correct the \textsc{noop} gate.

Consider, specifically, a control sequence composed of $N$ gates
$Q_{i}$ applied on $S$ in the presence of $H_{e}$. Ideally, let the
sequence be intended to implement a gate $A=Q_{N}\cdots Q_{1}$.  The
EPG $\Phi_{Q_{i}}$ associated with $Q_{i}$ is assumed to be an
arbitrary function of the gate $Q_{i}$: \[
\Phi_{Q_{i}}=\mathcal{E}(Q_{i}).\] $\mathcal{E}$ depends in principle
on $H_{e}$ as well as the details of implementation. Suppose, however,
that as in EDD no information about such dependence is directly or
indirectly incorporated into control design. We have the following:

\vspace*{1mm}

\textbf{Theorem.} \emph{Consider a sequence of gates
$A=Q_{N}Q_{N-1}\cdots Q_{1}$, and suppose that the combined
first-order error $\Phi_{A}$ for this sequence is equal to zero mod$B$
as long as the individual errors $\Phi_{Q_{i}}$ belong to a subspace
$\Omega_{e}\subseteq\Omega_{c}$ of dynamically correctable operators
for all $i$. If no further assumptions are made on $\Phi_{Q_{i}}$,
then the expectation values of operators in $\Omega_{e}$ are preserved
by $A$.}

\vspace*{1mm}

\textbf{Proof.} Since no assumptions are made on $\Phi_{Q_{i}}$ (hence
on $\Phi_{Q_{i}}$ mod$B$), the function $\mathcal{E}$ from unitary
system operators into $\Omega_{e}$ is arbitrary. In particular, we
may consider two error models defined by
\begin{eqnarray*}
\mathcal{E}_{1}(U)=X,\hspace{5mm}\mathcal{E}_{2}(U)=U^{\dagger}XU,
\end{eqnarray*}
\noindent
where $X$ is any operator in $\Omega_{e}$. Let $\Phi_{A,1}$
($\Phi_{A,2}$) denote the combined error obtained under the error
function $\mathcal{E}_{1}$ ($\mathcal{E}_{2}$). Let us define
$P_{i}=Q_{i}\cdots Q_{1}$ for $N\ge i>0$ and $P_{0}=I_{S}$. From
Eq. (\ref{eq:PhiA1}) and using the fact that
$Q_{i}=P_{i}P_{i-1}^{\dagger}$, we have:
\begin{eqnarray}
\Phi_{A,1}^{[1]} & = & \sum_{i=1}^{N}P_{i-1}^{\dagger}XP_{i-1}=0\text{
mod}B,\label{eq:vxv1}\\ \Phi_{A,2}^{[1]} & = &
\sum_{i=1}^{N}P_{i-1}^{\dagger}P_{i-1}P_{i}^{\dagger}
XP_{i}P_{i-1}^{\dagger}P_{i-1}\nonumber
\\ & = & \sum_{i=1}^{N}P_{i}^{\dagger}XP_{i}=0\text{
mod}B.\label{eq:vxv2}
\end{eqnarray}
\noindent
Comparing Eq. (\ref{eq:vxv1}) and Eq. (\ref{eq:vxv2}) yields:
\begin{eqnarray*}
\forall X\in\Omega_{e}: P_{0}^{\dagger}XP_{0}-P_{N}^{\dagger}XP_{N}
=  X-A^{\dagger}XA=0\text{ mod}B.
\end{eqnarray*}
\noindent We can write $X$ as $X=\tilde{X}+I_{S}\otimes B_{X}$ where
$\text{Tr}_{S}\tilde{X}=0$ so that $\tilde{X}$ has no pure-bath
component. The operator $A$ acts on the system only, thus we have
\begin{eqnarray*}
X-(A\otimes I_{B})^{\dagger}X(A\otimes I_{B})=0\text{ mod}B &
\Rightarrow & \!\!\!\tilde{X}-A^{\dagger}\tilde{X}A=0\\ & \Rightarrow
& \!\!\!  X-A^{\dagger}XA=0
\end{eqnarray*}
Clearly, $A\otimes I_{B}$ acts as identity on $\Omega_{e}$ and can
generate no non-trivial dynamics on this
subspace.\hfill{}$\blacksquare$

\vspace*{1mm}

In other words, we may yet effect a control-oblivious unitary
transformation on the system but it will necessarily commute with the
subspace of errors that ${\cal G}$ can correct for, and will thus
coincide with identity (\textsc{noop}) if the latter acts irreducibly
on ${\cal H}_{S}$.  Nonetheless, one might be interested in
suppressing errors from a particular \emph{subset} of more general
errors. The No-Go theorem then still allows such a black-box solution
to be viable for constructing primitive gates, as well as it allows to
suitably combining DD with quantum encoding. For examples of
constructions where decoherence is suppressed using BB pulses and the
desired computational dynamics is encoded in a subspace or subsystem
see \cite{Viola2000Dygen,Viola2002,Khodjasteh-Hybrid,lidar:160506}.

From a practical standpoint, the No-Go theorem places restrictions on
the ability to making fully portable DCGs where a fixed sequence of
arbitrary implementations of gates effects a generic unitary gate on
the system. In order to go beyond \textsc{noop} in constructing DCGs,
we thus require to include certain specified gate implementations that
circumvent the arbitrary EPG assumption.

\section{Dynamically Corrected Gates beyond NOOP}
\label{sec:dcgbeyond}

A central point in EDD is that the combined EPG is determined by the
variation of the control propagator between the gates at the vertexes
of the Cayley graph underlying the scheme. In view of the above No-Go
Theorem, identifying a procedure where an analogous error cancellation
is achieved but a non-trivial target unitary is effected on $S$,
requires \emph{access to relationships between errors associated to
different gates}. Identifying and exploiting such relationship is
possible because both the algebraic structure of $H_{e}$ and the
gating Hamiltonian are known. Thus, while the No-Go does not pose a
fundamental obstacle for DQEC, it does complicate the design of robust
control schemes and reduces their portability. It is the goal of this
Section to provide explicit procedures for constructing and analyzing
DCGs, by first illustrating the general approach and obtaining
estimates of the uncorrected (higher than first-order) net error, and
next providing full detail on linear decoherence- and dephasing-
protected DCGs. Both for added clarity, and because the presence of a
non-trivial $H_{S}\ne0$ requires a system-dependent analysis, we take
advantage in what follows of assumption \texttt{(a2)}, that is, we
explicitly let
\begin{equation}
H_{S}=0,\;\; H_{e}=H_{SB}+H_{B},\;\;
H_{\text{gate}}(t)=H_{\text{ctrl}}(t),\label{driftless}
\end{equation}
\noindent
and further discuss the role of $H_{S}$ in DQEC constructions in
Sec. \ref{sub:extenapp}.

\subsection{DCG constructions and error estimates}
\label{sub:dcgconstruction}

Probably the simplest (albeit not unique) way to enforce the
constraint of non-trivial gate error relationships is to imagine that
two different gates share the same EPG to lowest order. Let $Q$ be the
intended (primitive) target gate, with associated error $\Phi_{Q}$,
and imagine that a special \textsc{noop} gate $I_{Q}$ is available,
such that the error associated with $I_{Q}$ is also $\Phi_{Q}$:
\begin{equation}
U_{Q}=Q\,\exp(-i\Phi_{Q}),\;\;\; I_{Q}=I_{S}\,\exp(-i\Phi_{Q}).
\label{same}
\end{equation}
\noindent
That the above requirement can in fact be fulfilled will be addressed
in Sec. \ref{sub:diffequal}. Assuming for now that Eq. (\ref{same})
holds, consider attaching a self-directed edge $I_{Q}$ to each vertex
on the Cayley graph $G(\mathcal{G},\Gamma)$, and let the Eulerian path
${\cal C}$ be modified so that the new edges added at each vertex are
incorporated. The error associated with this new sequence (which
clearly still acts as \textsc{noop}) is given, up to the first order,
by \[
\Phi_{\text{DCG}}^{[1]}=\Phi_{\text{EDD}}^{[1]}+\Pi_{\mathcal{G}}(\Phi_{Q}),\]
and is canceled (mod$B$) as long as the errors $\Phi_{Q}$ and
$\Phi_{F_{j}}$ belong to $\Omega_{e}$. Next, we change the final
gate/edge $I_{Q}$ in the path we just created with the gate/edge $Q$
with error $\Phi_{Q}$.  By construction, the error associated with the
combined sequence remains unchanged, but the sequence now implements
$Q$ as opposed to $I_{S}$.  Thus, we have succeeded in obtaining a DCG
that performs the desired gate $Q$ on $S$ with a smaller error as
compared to its uncorrected implementation.

In line with the philosophy underlying DQEC approaches, the errors due
to $H_{e}$ are \emph{suppressed asymptotically}. This requires that
the EPG of the primitive (uncorrected) gates be small to begin with in
order to expect a relative improvement in DCG performance {[}recall
Eq. (\ref{dcgl}){]}. Even if this condition is met, the uncorrected
errors associated with a large number of operations performed in
succession tend to inevitably build up, thus additional precautions
must be taken to avoid catastrophic error growth. Within FTQEC, this
can still be prevented provided that DCG residual errors remain
correctable by the given code architecture and smaller than the
relevant threshold.  This prompts us to estimate the uncorrected
errors associated with DQEC constructions for each gate.

The asymptotic nature of DCG constructions rests on the validity of
the Magnus expansion as an approximation of the effective Hamiltonian
for short time intervals. When used to estimate the EPG associated
with a gate $Q$ of duration $T=N\tau$, a \emph{sufficient} condition
is \begin{equation} \norm{H_{e}}T<\pi,
\label{eq:magncond}
\end{equation}
\noindent
where $N=1$ corresponds to a single primitive gate segment. Two
remarks are in order. First, in view of assumption \texttt{(c2)},
$\tau_{\text{min}}$ effectively limits the strength of error
Hamiltonians that can be corrected using DCGs. Second, while including
the environment Hamiltonian $H_{B}$ in $H_{e}$ has the advantage of
simplicity, the resulting convergence radius (time) for the Magnus
expansion tends to be overly pessimistic in this way, in particular it
may grossly overestimate the error in the case where
$\norm{H_{B}}\gg\norm{H_{SB}}$.  One solution is to move to a toggling
frame that removes the evolution of $B$. We refer for instance to
\cite{Khodjasteh2005} for an analysis along this lines, yielding a
tighter convergence radius for the Magnus expansion,
$\norm{H_{SB}}T<\pi$.

The uncorrected errors that do \emph{not} cancel as a result of a DQEC
protocol can be bounded by the second- and higher- order contributions
to the Magnus series \cite{Note4}. Recalling Eq. (\ref{eq:Het}), we
can estimate the uncorrected error $\Phi_{Q,\text{DCG}}$ associated
with a DCG $Q$ as
\begin{eqnarray} \norm{\Phi_{Q,\text{DCG}}^{[2+]}} &
\!=\! &
\frac{1}{2}\norm{\int_{0}^{T}
\int_{0}^{t_{2}}[H_{e}(t_{1}),H_{e}(t_{2})]dt_{1}dt_{2}}\nonumber
\\ & \!\le\! &
\frac{T^{2}}{4}\max_{t_{1}<t_{2}}\norm{\,[H_{e}(t_{1}),H_{e}(t_{1})]\,}
\label{eq:undecerr}\\
& \!\le\! &
\frac{T^{2}}{4}\Big(2\,\norm{H_{B}}\norm{H_{SB}}+\norm{H_{SB}}^{2}\Big),
\nonumber
\end{eqnarray}
\noindent
where we have used the unitary invariance properties of
$\norm{\cdot}$.  While Eq. (\ref{eq:undecerr}) gives an upper bound
for the uncorrected error associated with any gate $Q$ for which
$\Phi_{Q}^{[1]}=0$ mod$B$, it fails to capture many interesting
features of the interplay between the pure environment evolution and
the system-bath coupling.  We will reconsider the latter numerically
in the specific setting of Sec. \ref{sec:sim}.

Note that the first order error $\Phi_{Q}^{[1]}$ associated with a
primitive gate $Q$ (not corrected through DQEC) will naturally depend
on the number of qubits $n$ in the system. For the linear decoherence
model we have \[ \norm{\Phi_{Q}^{[1]}}\propto n,\] thus the EPG
\emph{per qubit} is expected to be constant (size-independent) up to
the first order. For DCGs, the first order error is zero, and to
obtain a dependence on the $n$, we need to focus on
$\Phi_{Q,\text{DCG}}^{[2]}$, estimated above. Using
Eq. (\ref{eq:undecerr}), we may write \[
\norm{\Phi_{Q,\text{DCG}}^{[2+]}}=O(n^{s}),\;\;\;1\le s\le2.\] The
upper bound is given by $\norm{H_{SB}}^{2}$
{[}Eq. (\ref{eq:undecerr}){]}, whereas the lower bound is achieved,
for example, when the bath operators $B_{\alpha}^{(i)}$ act
\emph{locally},
$[B_{\alpha}^{(i)},B_{\beta}^{(j)}]\propto\delta_{i,j}$, for all
$\alpha,\beta$.

\subsection{Finding gates with same error}
\label{sub:diffequal}

As described in Sec. \ref{sub:dcgconstruction}, a fundamental
ingredient in DCGs is that two control sequences are found for every
desired gate $Q$: One designated to perform a \textsc{noop} and the
other designated to perform $Q$, such that the corresponding errors
are equal up to the first order. We now give a simple method for
generating such sequences, which in the language of
Ref. \onlinecite{khodjasteh-2009} realize a so-called {\em first-order
balance pair}.

Suppose that a given control input $h_{\alpha}^{Q}(t)$ is intended to
produce a unitary gate $Q$ during an interval $[0,\tau]$, and let
$H_{\text{gate}}^{Q}$, and $U_{\text{gate}}(t)$ denote the gating
Hamiltonian and corresponding propagator for $Q$ over $t\in[0,\tau]$,
respectively. The error associated with $Q$ is given by
\begin{eqnarray*}
\exp(-i\Phi_{Q}) & \!=\! & T_{+}\Big[\exp\Big(\!-i\int_{0}^{\tau}\!
H_{e}(s,0)ds\Big)\Big],
\end{eqnarray*}
with $H_{e}(t,0)$ being the toggling frame error Hamiltonian as in
Eq. (\ref{eq:Het}). We next smoothly extend the control profile for
$Q$ to $[0,2\tau]$ by applying a gate $Q'$ immediately after $Q$, in
such a way that the overall control propagator for $t\in[0,2\tau]$ is
given by:
\begin{eqnarray*}
U'_{\text{gate}}(t,0)=\left\{ \begin{array}{ll}
U_{\text{gate}}(t), & 0<t<\tau;\text{ gate }Q, \\
U_{\text{gate}}(2\tau-t), & \tau<t<2\tau;\text{ gate }Q'.
\end{array}\right.
\end{eqnarray*}
The extended (composite) gate $I_{Q}=Q'Q$ implements the identity
(\textsc{noop}), with a first-order error given by:
\begin{eqnarray}
\Phi_{Q'Q}^{[1]} & = &
\int_{0}^{2\tau}U'_{\text{gate}}(t,0)^{\dagger}H_{e}U'_{\text{gate}}(t,0)dt
\label{eq:phiqq'}\\
& = &
2\int_{0}^{\tau}U_{\text{gate}}(t)^{\dagger}H_{e}U_{\text{gate}}(t)dt=2\,
\Phi_{Q}^{[1]}.\nonumber
\end{eqnarray}

Consider next another control propagator associated with a gate
$Q_{1/2}$ over the interval $[0,2\tau]$ obtained by letting \[
U''_{\text{gate}}(t,0)=U_{\text{gate}}(t/2),\;\;\;0<t<2\tau.\]
\noindent
The first-order error associated with $Q_{1/2}$ reads
\begin{eqnarray}
\Phi_{Q_{1/2}}^{[1]} & = &
\int_{0}^{2\tau}U''_{\text{gate}}(t,0)^{\dagger}H_{e}
U''_{\text{gate}}(t,0)dt\label{eq:phiq12}\\
& = &
2\int_{0}^{\tau}U_{\text{gate}}(t)^{\dagger}H_{e}
U_{\text{gate}}(t)dt=2\,\Phi_{Q}^{[1]}.\nonumber
\end{eqnarray}
\noindent
By comparing Eqs. (\ref{eq:phiqq'})-(\ref{eq:phiq12}), we observe that
while $Q'Q$ ideally implements the identity and $Q_{1/2}$ implements
$Q$, the EPGs resulting from $H_e\ne 0$ are equal up to the first
order.

\begin{figure}[tb]
\begin{centering}
\includegraphics[width=2in]{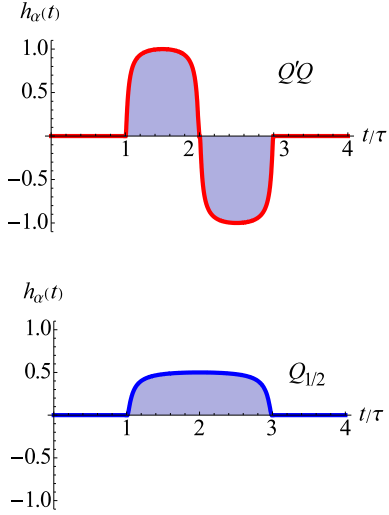}
\par\end{centering}
\caption{(color online) Control profiles for implementing two gates
$I_Q = Q'Q$ and $Q_{1/2}$, that correspond to a target identity gate
and a target gate $Q$, with equal first-order error.
\label{fig:profile}}
\end{figure}

Given enough control over the $\{h_{\alpha}(t)\}$, simple recipes may
be given for implementing both $Q'$ and $Q_{1/2}$. For instance, $Q'$
may be realized by applying $-h_{\alpha}^{Q}(t-\tau)$ (the reverse
anti-symmetric profile to $h_{\alpha}^{Q}(t)$) when $t>\tau$, see
Fig. \ref{fig:profile}.
%
Likewise, to implement $Q_{1/2}$, we may use a control input
$h_{\alpha}^{Q_{1/2}}(t)$ such that
$\int_{0}^{t}h_{\alpha}^{Q_{1/2}}(s)ds=\int_{0}^{t/2}h_{\alpha}^{Q}(s)ds$.
For example, if the pulse profiles are rectangular, $Q_{1/2}$ may be
generated by using a control input with half the speed and half the
strength of the one used for $Q$. The two prescriptions so obtained
for $QQ'$ and $Q_{1/2}$ corresponds then, respectively, to the
piecewise-constant profiles given in Eqs. (4)-(5) of
\cite{khodjasteh-2008}. If we assume that all primitive gates have the
\emph{same duration} $\tau$, then the number of primitive gates
required in each DCG will be $dm+2d=d(m+2)$ {[}recall
Sec. \ref{sub:dcgconstruction}{]}.

The construction described in this section should, in its simplicity,
mainly be taken as a proof-of-concept illustration of how to meet the
fundamental `same error requirement' of Eq. (\ref{same}). In practice,
numerical optimization techniques may prove either helpful or
indispensable, especially in the presence of additional control-input
constraints.  For example, if negative control inputs are not
available to implement the required balance pair $(I_Q,Q_*)$ as
$(Q'Q,Q_{1/2})$, we can invoke a more general construction given in
\cite{khodjasteh-2009}.  The latter only assumes access to stretchable
control profiles to implement $(I_Q,Q_*)$, with a resulting sequence
that is longer by a factor of $3/2$ with respect to $(Q'Q,Q_{1/2})$.
The important point, however, is that any DCG construction
necessitates a recipe for the control inputs that will logically
produce a non-trivial relationship among the corresponding EPGs, in
order to evade the No-Go theorem of Sec. \ref{subsec:contobv}.

\subsection{Permissible controls for general linear decoherence and
pure dephasing}
\label{sec:Examples}

We next specialize the above general constructions to two physically
relevant error models on $n$ (driftless) qubits: (i) the generic
linear decoherence model described by Eq. (\ref{eq:Hsbsingle}), where
$\Omega_{e}=\Omega_{e}^{\{1\}}$ includes arbitrary single-qubit error
operators, Eq. (\ref{eq:Hsbsingle1}); (ii) the pure dephasing (or
single-axis decoherence) model, which is formally obtained from the
above generic case by assuming that coupling to $B$ occurs along a
known direction, say $Z$, in which case $B_{\alpha}^{(i)}=0$ for
$\alpha=1,2$. We will refer to the resulting subspace of system-bath
error operators,
\begin{equation}
\Omega_{e}=\text{span}\{Z^{(i)}\otimes
B_{z}^{(i)}\}\equiv\Omega_{e}^{Z},
\end{equation}
\noindent
as the `(pure) phase errors'. The dephasing model is a common
approximation often used in situations where a quantization axis
defines the internal states of the individual qubits (such as in
standard NMR QIP settings \cite{Cory-Overview1}), and allows for
significantly less resource-intensive FTQEC schemes to be devised in
comparison to general linear decoherence \cite{aliferis:08}. This is
not only due to the fact that the errors involved in dephasing are
inherently restricted, but also to the fact that the algebraic
structures associated with their growth are much simpler \cite{Note5}.

\subsubsection{One- and two-body controllable Hamiltonians}
\label{sub:contham}

Within the network model of QC, the majority of proposals invoke the
use of control Hamiltonians that act non-trivially on single or pairs
of qubits. Let us consider the set of primitive gates which are
generated by switching on and off control input in a control
Hamiltonian parametrized as follows:
\begin{eqnarray}
H_{\text{ctrl}}(t) & = &
H_{\text{ctrl}}^{(\text{pair})}(t)+H_{\text{ctrl}}^{(\text{single})}(t),
\label{onetwo}\\
H_{\text{ctrl}}^{(\text{single})}(t) & = &
\sum_{i=1}^{n}\sum_{\alpha=X,Y}h_{\alpha}^{(i)}(t)S_{\alpha}^{(i)},\nonumber
\\ H_{\text{ctrl}}^{(\text{pair})}(t) & = & \sum_{i\ne
j}\sum_{\alpha=1}^{3}h_{\alpha\alpha}^{(ij)}(t)S_{\alpha}^{(i)}
S_{\alpha}^{(j)},\nonumber
\end{eqnarray}
\noindent
where we generally allow for homogeneous two-body interactions which
include natural entangling Hamiltonians such as the Ising $(ZZ)$ and
Heisenberg $(\mathbf{S}\cdot\mathbf{S}\equiv XX+YY+ZZ)$ interactions.
We will focus on primitive gates during which the control propagator
can be expanded as
\begin{eqnarray} U_{\text{ctrl}}(t,0) & = &
\exp\left[-iA^{\{1\}}(t)-iA^{\{2\}}(t)\right],
\label{eq:a1atu}\\
A^{\{1\}}(t) & = &
\sum_{i=1}^{n}\sum_{\alpha=1}^{3}\theta_{\alpha}^{(i)}(t)
S_{\alpha}^{(i)},\nonumber
\\ A^{\{2\}}(t) & = &
\sum_{i<j}\sum_{\alpha=1}^{3}\theta_{\alpha}^{(i,j)}(t)
S_{\alpha}^{(i)}S_{\alpha}^{(j)},\nonumber
\end{eqnarray}
\noindent
where the functions $\theta_{\alpha}^{(i)}(t)$,
$\theta_{\alpha}^{(i,j)}(t)$ are suitable control integrals and the
set of qubits involved in two-qubit control has \emph{no} intersection
with the set of qubits involved in single-qubit control. That is, we
require that
\begin{equation}
\{i|\theta_{\alpha}^{(i)}(t)\ne0\}\cap\{j|
\theta_{\alpha}^{(k,j)}(t)\ne0\}=\emptyset,
\label{eq:divided}
\end{equation}
\noindent
so that $A_{\{1\}}(t)$ and $A_{\{2\}}(t)$ commute. For example,
applying a $ZZ$ Hamiltonian between qubits 1 and 2, while qubits 3, 4,
and 5 are affected by single-qubit Hamiltonians is an example where
the above constraint is naturally obeyed. Note that such a `divided
control' is implicit in the circuit model of QC where gates are
applied in parallel only when they commute. We emphasize that unless
certain commutativity conditions are satisfied, in general
$\theta_{\alpha}^{(ij)}(t)\ne\int_{0}^{t}h_{\alpha}^{(ij)}(s)ds$, and
a linear relationship among the control inputs does not translate into
a linear relationship among the control integrals.

\subsubsection{Error-corrected gate constructions}
\label{sub:decsec}

\begin{figure}[tb]
\begin{centering}
\includegraphics[width=3.3in]{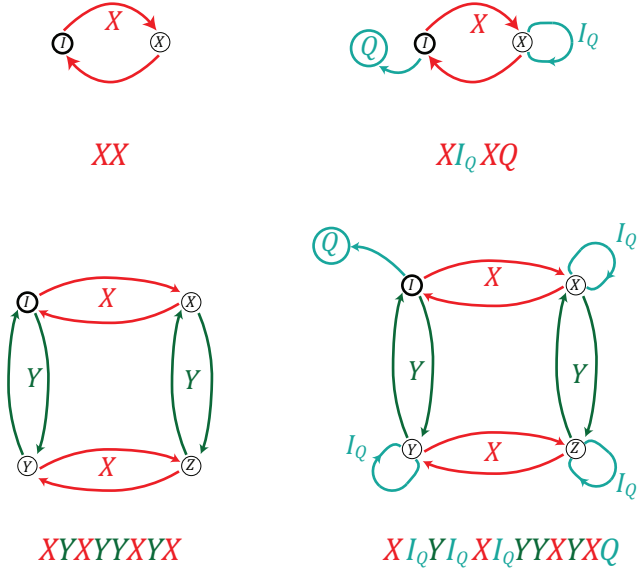}
\par\end{centering}
\caption{(color online) Left: Cayley graphs for $\mathbb{Z}_{2}$ (top)
and $\mathbb{Z}_{2}\otimes\mathbb{Z}_{2}$ (bottom), relevant to pure
dephasing and arbitrary linear decoherence,
respectively. Representative Eulerian cycles and the corresponding EDD
sequences are depicted.  Right: Modified Cayley graphs for DCG
constructions based on $\mathbb{Z}_{2}$ (top) and
$\mathbb{Z}_{2}\otimes\mathbb{Z}_{2}$ (bottom), respectively.
Relevant Eulerian paths and the corresponding DCG sequences are
depicted.
\label{fig:cayleys}}
\centering{}
\end{figure}

A good (minimal) DD group for the error set $\Omega_{e}^{\{1\}}$
corresponding to arbitrary linear decoherence is given by \[
\Omega_{e}^{\{1\}}\rightarrow\mathcal{G}_{\text{LD}}=
\mathbb{Z}_{2}\otimes\mathbb{Z}_{2},\] represented in ${\cal
H}_{S}=({\mathbb{C}}^{2})^{\otimes n}$ through the $n$-fold tensor
power representation
$\{S_{\alpha}^{(\text{all})}\}=\{I_{S},X^{(\text{all})},
Y^{(\text{all})},Z^{(\text{all})}\}$
\cite{Viola1999Dec,Viola2003Euler}. $\mathcal{G}_{\text{LD}}$ has two
generators, which under the above representation we may take to be
$\{F_{j}\}=\{X^{(\text{all})},Y^{(\text{all})}\}$. Observe that the
subspace of operators decoupled by $\mathcal{G}_{\text{LD}}$ is not
limited to $\Omega_{e}^{\{1\}}$, as it includes, for example,
inhomogeneous bilinear couplings of the form
$S_{\alpha}^{(i)}S_{\beta}^{(j)}$, with $i\ne j$ and
$\alpha\ne\beta$. Thus,
$\Omega_{c,\text{LD}}\supset\Omega_{e}^{\{1\}}$, which allows a wider
class of primitive gates to be corrected using
$\mathcal{G}_{\text{LD}}$ than the \textsc{noop} gate with
single-qubit errors as originally analyzed in \cite{Viola2003Euler}.

The DD group for pure phase errors is simpler: \[
\Omega_{e}^{Z}\rightarrow\mathcal{G}_{Z}=\mathbb{Z}_{2},\] represented
via $\{I_{S},X^{(\text{all})}\}$, and with the single generator
$\{F\}=\{X^{(\text{all})}\}$. Notice again that this representation
also decouples errors beyond the simple dephasing terms, for instance
inhomogeneous error operators of the form $Z^{(i)}X^{(j)}$ with $i\ne
j$, therefore again we have
$\Omega_{c,\text{Z}}\supset\Omega_{e}^{\{Z\}}$.  A pictorial view of
the Cayley graphs relevant to both error models under examination is
given in the left panels of Fig. \ref{fig:cayleys}, along with
representative Eulerian cycles.

A possible EDD sequence for correcting \textsc{noop} under the general
decoherence model is given by: \begin{equation}
\text{EDD}^{\text{lin}}\rightarrow
X^{(\text{all})}Y^{(\text{all})}X^{(\text{all})}Y^{(\text{all})}
Y^{(\text{all})}X^{(\text{all})}Y^{(\text{all})}X^{(\text{all})},
\label{eq:eddge}
\end{equation}
\noindent
where operations are understood to be applied from right to left,
following the underlying Eulerian cycle. This implies a total number
of $md=4(\text{group elements})(2\text{ generators})=8$
gates. Similarly, EDD under pure dephasing may be implemented by the
sequence:
\begin{equation}
\text{EDD}^{\text{Z}}\rightarrow
X^{(\text{all})}X^{(\text{all})},
\end{equation}
\noindent
which consists of just $md=2$ gates, and may be viewed as implementing
a form of `continuous-time' spin-echo \cite{Viola2004,Nir}.  The
relevant DCG constructions are based on modifications of the above EDD
gate sequences. Let $Q$ denote a primitive gate. We define $I_{Q}=QQ'$
and $Q_{1/2}$ using the constructions of Sec. \ref{sub:diffequal}. For
general linear decoherence, a DCG for $Q$ is then given by:
\begin{widetext}
\begin{equation}
\text{DCG}^{\text{lin}}\rightarrow
Q_{1/2}X^{(\text{all})}Y^{(\text{all})}X^{(\text{all})}Y^{(\text{all})}
I_{Q}Y^{(\text{all})}I_{Q}X^{(\text{all})}I_{Q}Y^{(\text{all})}
I_{Q}X^{(\text{all})}.
\label{eq:dcgld}
\end{equation}
\end{widetext}
\noindent
Similarly, for pure dephasing, we simply have:
\begin{equation}
\text{DCG}^{\text{Z}}\rightarrow
Q_{1/2}X^{(\text{all})}I_{Q}X^{(\text{all})}.
\label{eq:dcgph}
\end{equation}
\noindent
The sequences given in Eqs. (\ref{eq:dcgld})-(\ref{eq:dcgph})
represent Eulerian paths in the modified Cayley graphs corresponding
the DD groups $\mathcal{G}_{\text{LD}}$ and $\mathcal{G}_{Z}$,
obtained as described in Sec. \ref{sub:dcgconstruction}; see also
right panels in Fig. \ref{fig:cayleys}. If all primitive gates have
the same duration $\tau$, since each added `arm' in the modified graph
has duration $2\tau$, the resulting time overheads may be summarized
as follows:

\vspace*{1mm}

\begin{center}
\begin{tabular}{c|cc}
 & $\;$\texttt{Linear decoherence}  & $\;\;$\texttt{Pure dephasing}
\tabularnewline
\hline
EDD$\;$  & 8  & 2 \tabularnewline
DCG$\:$  & 16  & 6 \tabularnewline
\hline
\end{tabular}
\par\end{center}

\subsubsection{Error per gate structure}
\label{sub:epgstruct}

In order for the DQEC constructions just provided to be valid, it is
necessary to explicitly show that the errors associated with the
primitive gates generated from the (divided) one- and two-body
Hamiltonians of Sec. \ref{sub:contham} can indeed be decoupled by the
appropriate DD group.

Consider linear decoherence and $\mathcal{G}_{\text{LD}}$ first.  Let
the control propagator for a primitive gate $Q$ be given by Eq.
(\ref{eq:a1atu}). The toggling frame error Hamiltonian $H_{e}(t)$
{[}Eq. (\ref{eq:Het}){]} is given by
\begin{eqnarray*} H_{e}(t) & = &
e^{i[A^{\{1\}}(t)+A^{\{2\}}(t)]}H_{e}e^{-i[A^{\{1\}}(t)+A^{\{2\}}(t)]}\\
& = &
e^{iA^{\{2\}}(t)}e^{iA^{\{1\}}(t)}H_{e}e^{-iA^{\{1\}}(t)}e^{-iA^{\{2\}}(t),}
\end{eqnarray*}
\noindent
where we used the fact that $[A^{\{1\}}(t),A^{\{2\}}(t)]=0$. Notice
next that
$\exp\left[iA^{\{1\}}(t)\right]H_{e}\exp\left[-iA^{\{1\}}(t)\right]$
can be expanded as a sum of {\em single-qubit terms} up to the first
order Magnus, thus
\begin{eqnarray*}
e^{iA^{\{1\}}(t)}H_{e}e^{-iA^{\{1\}}(t)}=\sum_{\alpha,i}S_{\alpha}^{(i)}\otimes
C_{\alpha}^{(i)}\equiv F_{e}(t)\in\Omega_{e}^{\{1\}}.
\end{eqnarray*}
\noindent
Up to the first order we may then write:
\begin{eqnarray*}
\Pi_{\mathcal{G}_{\text{LD}}}[H_{e}(t)]=e^{iA^{\{2\}}(t)}
\Pi_{\mathcal{G}_{\text{LD}}}[F_{e}(t)]e^{-iA^{\{2\}}(t)}=0\mbox{
mod$B$,}
\end{eqnarray*}
\noindent
where we used the fact that
$[e^{iA^{\{2\}}(t)},S_{\alpha}^{(\text{all})}]=0$.  Subsequently, the
first-order error $\Phi_{Q}^{[1]}$ is decoupled by
$\mathcal{G}_{\text{LD}}$: \[
\Pi_{\mathcal{G}_{\text{LD}}}(\Phi_{Q}^{[1]})=
\Pi_{\mathcal{G}_{\text{LD}}}\Big(\int_{0}^{t}H_{e}(s)ds\Big)=0\mbox{
mod$B$,}\] which establishes the desired error cancellation.

For the pure dephasing model, the set of allowed control Hamiltonians
in DCGs is slightly more restricted. For example, consider a DCG
construction based on Eq. (\ref{eq:dcgph}), where $\mathcal{G}_{Z}$ is
represented by $\{I,X^{(\text{all})}\}$. In this case, the
single-qubit control Hamiltonians used in $Q$ cannot include $Y^{(i)}$
terms as they would result in errors that are not decoupled by this
particular representation.  However, one can alternatively use the
representation $\{I,Y^{(\text{all})}\}$ of $\mathcal{G}_{Z}$, if the
control Hamiltonians used in $Q$ are given by $Y^{(i)}$. The choice of
the particular representation of $\mathcal{G}$ used in constructing
DCGs may thus affect the set of permissible Hamiltonians. While this
is not surprising in view of existing results on control of decoupled
evolutions \cite{Viola1999Control}, analyzing in full generality the
compatibility of primitive gate sets with DQEC strategies for a given
error model of interest warrants a separate investigation.

\section{Case study: Dynamical correction of spin-bath decoherence}
\label{sec:sim}

In this section, we specialize the system-independent analytic DCG
constructions described in Sec. \ref{sec:dcgbeyond} to the
paradigmatic setting of spin-bath decoherence. While quantitative
modeling of a specific device is not our purpose here, our analysis is
inspired by spin-based QIP architectures -- in particular, electron
spin qubits in semiconductor quantum dots, as considered for instance
in Refs. \onlinecite{Loss,HansonRev,WenReview}.  In which physical
parameter regime(s) is the improvement predicted for DQEC methods
actually to be seen?  What distinctive physical features and
performance trade-offs are associated with DCG constructions?  Beside
complementing and expanding our previous numerical results
\cite{khodjasteh-2008}, these are two questions that the present
investigation further addresses by example.

\subsection{Model system and primitive gates\label{sub:modelsystem}}

Let the system $S$ consist of $n$ individually addressable spin-$1/2$
degrees of freedom, and let the environment $B$ likewise consist of
$n_{B}$ spin-$1/2$ particles. If we denote by $\mathbf{I}^{(i)}$ the
spin vector operator associated with the $i$-th bath particle, the
internal Hamiltonian $H_{\text{int}}=H_{B}+H_{SB}$ we consider reads:
\begin{eqnarray*}
H_{B} & = & \sum_{1\le i<j\le
n_{B}}\Gamma_{i,j}(\mathbf{I}^{(i)}\cdot
\mathbf{I}^{(j)}-3I_{Z}^{(i)}I_{Z}^{(j)}),\\
H_{SB} & = &
\sum_{i=1}^{n}\sum_{k=1}^{n_{B}}A_{k}^{(i)}
\mathbf{S}^{(i)}\cdot\mathbf{I}^{(k)}.
\end{eqnarray*}
\noindent
Physically, $H_{B}$ describes dipolar interactions between the bath
spins, whereas $H_{SB}$ describes a (hyperfine) contact interaction of
each qubit with the bath. For concreteness, we further assume in what
follows that the coupling strength $\Gamma_{i,j}$ between bath spins
$i$ and $j$ is arbitrarily chosen from a uniform random distribution
in $[-\Gamma,\Gamma]$, and similarly that the hyperfine coupling
strengths $A_{k}^{(i)}$ for each qubit are independently sampled
uniformly at random from the $[-A,A]$ interval.

In line with Eq. (\ref{onetwo}), control over individual qubits and
pairs of qubits is introduced through a gating Hamiltonian of the form
\[
H_{\text{gate}}(t)=\sum_{i=1}^{n}\sum_{\alpha=X,Y}h_{\alpha}^{(i)}(t)
S_{\alpha}^{(i)}+\sum_{i,j=1}^{n}h_{W}^{(i,j)}(t)\mathbf{S}^{(i)}
\cdot\mathbf{S}^{(j)},\]
\noindent
where $\mathbf{S}^{(i)}\cdot\mathbf{S}^{(j)}$ is the Heisenberg
exchange interaction \cite{Note6}. Thus, each primitive gate is
realized by switching the control parameters $h_{X}^{(i)}(t)$,
$h_{Y}^{(i)}(t)$, and $h_{W}^{(i,j)}(t)$, under the `divided control'
assumption of Eq. (\ref{eq:divided}). In particular, we can simply
assume that at each time $t$ the qubits controlled through $X^{(i)}$,
$Y^{(i)}$, and $\mathbf{S}^{(i)}\cdot\mathbf{S}^{(j)}$ are distinct,
and use as a starting point for DQEC the universal set of primitive
gates given by ideal gates $\{\exp(-i\theta C)\}$, where $C$ can be
$X^{(i)}$, $Y^{(i)}$, and $\mathbf{S}^{(i)}\cdot\mathbf{S}^{(j)}$. We
restrict available control inputs to the simplest choice of
rectangular profiles: thus, a primitive gate is simply achieved by
turning on the corresponding control input for a duration $\tau$ at a
fixed strength $h$, allowing any control Hamiltonian to be written as
a piece-wise constant function of time. The choice of the rectangular
profiles allows for numerically exact simulations \cite{Note7}, while
avoiding inessential complications in describing relevant control
constraints and imperfections. In particular, the bounded-strength
control constraint \texttt{(c1)} translates into demanding that
$\text{max}_{t}\abs{h_{Q}(t)}<h_{\text{max}}$, while the constraint
\texttt{(c2)} implies a shortest gating time $\tau \geq
\tau_{\text{min}}$ {[}Sec. \ref{sub:contobj}{]}. As an illustrative
example of control errors, we may define over-rotation errors by
replacing the nominal rotation angle $2\theta$ in each (piece-wise
constant) control segment with $2\theta(1+\varepsilon)$. If
$\varepsilon$ fluctuates non-deterministically over different
instances of a particular gate, we have a random over-rotation; a
constant (realization-independent) $\varepsilon$ defines a systematic
over-rotation.

\par Within the above setting, the internal Hamiltonian is the error
Hamiltonian, $H_{e}=H_{\text{int}}$ and, as remarked, there is no
drift, $H_{S}=0$.  Since $H_{SB}$ belongs to $\Omega_{e}^{\{1\}}$, as
described in Sec. \ref{sec:DDNOOP} the relevant DD group
$\mathcal{G}_{\text{LD}}$ discussed in \ref{sub:decsec}. Note that the
DD problem for a single electron-spin qubit in contact with a
nuclear-spin bath has been extensively analyzed using BB pulses in
\cite{Wen01,Wen02,Wen03}, the limit $H_{S}\approx0$ corresponding to
the low-bias regime where transverse and longitudinal relaxation fully
compete. In our case, the two available group generators are realized
through collective spin-flip gates, and the DCG constructions of
Sec. \ref{sub:dcgconstruction} may be used. Thus, a primitive gate
$Q=\exp(-i\theta C)$ of duration $\tau$ is converted to a DCG gate
through the following control Hamiltonian:
\begin{equation}
H_{\text{DCG}}(t)=\begin{cases} H_{i}, & t\in[(i-1)\tau,i\tau]\text{,
}i=1,\cdots,16,\\ 0 &
t\notin[0,16\tau],\end{cases}\label{eq:hdcgt}\end{equation} where
explicitly we have (cf. also the DCG circuit given in Fig.  1 of
\cite{khodjasteh-2008}): \[ H_{i}=\frac{1}{\tau}\begin{cases}
\frac{\pi}{2}\left(X^{(1)}+\cdots+X^{(n)}\right),\, & i=1,7,12,14,\\
\frac{\pi}{2}\left(Y^{(1)}+\cdots+Y^{(n)}\right),\, & i=4,10,11,13,\\
+\theta C, & i=2,5,8,\\ -\theta C, & i=3,6,9,\\ +\frac{\theta}{2}C, &
i=15,16.\end{cases}\] Notice that besides the original primitive gate
$Q$ used in segments $i=2,5,8$, in segments $i=3,6,9$ we implement the
gate $Q^{-1}$ (needed for the special \textsc{noop} gate $I_{Q}$)
through negative control inputs, while $Q_{1/2}$ in segments $i=15,16$
is obtained by using half the control power than for $Q$.

\subsection{Numerical results}
\label{sub:num}

Perhaps the most important and obvious question, if DCGs are to be
incorporated in a QC architecture, is whether they can usefully
improve EPGs. While in theory DCG constructions are provably effective
in the asymptotic limit of small errors, how stringent in practice
this limit might be, can sensitively depend upon the specifications
and operating constraints of the device technology at hand. For this
reason, quantitative predictions on the effectiveness of DCGs are thus
best formulated and addressed having a specific experimental platform
in mind. Numerical simulations in toy models as we examine can yet be
instrumental in developing intuition on how to map out the actual
regime of improvement as different parameters are varied -- in
preparation for realistic scenarios where both $H_{SB}$ and $H_{B}$
are typically fixed by either physical or fabrication constraints.

In the numerical analysis that follows, we explore a wide range of
open-system as well as control parameters, specifically $A$, $\Gamma$,
and the base gate interval $\tau$, which for the current purpose can
be also thought as identifying the shortest accessible modulation
timescale, $\tau\equiv\tau_{\text{min}}$. In order to allow for
arbitrary precision matrix calculations (see also the Appendix), we
focus on a relatively small open system, letting $n=2$, $n_{B}=6$
henceforth. In all simulations, the qubit register is initialized in
the fixed reference state
$|\psi_{\text{in}}\rangle=({\ket{00}+\ket{01}}/{\sqrt{2}})$, whereas
the environment is initially maximally mixed,
$\rho_{B}^{\text{in}}=I_{B}/2^{n_{B}}$.  The evolution of the combined
density matrix for the system and the environment is then calculated
by multiplying exact matrix exponentiation for piecewise constat
Hamiltonians describing the segments of the evolution.  As
representative target quantum gates, we consider single- and two-qubit
rotations by $\pi/4$, that is:
\begin{eqnarray*}
R_{\pi/4}^{(1)}=\exp[-i\,({\pi}/{8})X^{(1)}]\otimes I^{(2)},\\
W_{\pi/4}^{(1,2)}=\exp[-i\,({\pi}/{8})\mathbf{S}^{(1)}
\cdot\mathbf{S}^{(2)}]\equiv\sqrt{\textsc{swap}},
\end{eqnarray*}
\noindent
where the last equality makes explicit contact with the (universal)
square-root-of-swap gate \cite{NielsenBook,Loss}. Both the
(first-order) improvement ratio introduced in
Eqs. (\ref{minfid})-(\ref{ratio}), $r^{[1]}\equiv r$, or directly the
corresponding (in)fidelity contributions for the given initial state,
will be used as a metric for quantifying DCG performance. %

\begin{figure}[t]

\begin{centering}
\includegraphics[width=3.2in]{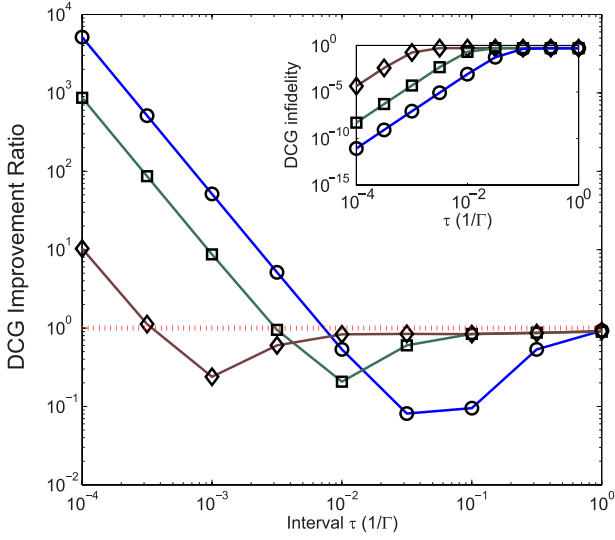}
\par\end{centering}

\centering{}
\caption{(color online) Improvement ratio as a function of (minimum)
gating interval $\tau$, for the two-qubit target gate
$W_{\pi/4}^{(1,2)}$.  The `no improvement' (horizontal) line $r=1$ is
included for reference in this and all other figures that use $r$ as a
metric for performance.  The internal-bath Hamiltonian is kept fixed
at $\Gamma=1$, whereas $A/\Gamma$ takes the value $10^{2}$ {[}brown
diamonds{]}, $10^{1}$ {[}green squares{]} and $1$ {[}blue
circles{]}. Inset: Infidelity error for the corresponding DCG
implementation, $1-f_{W_{\text{DCG}}}$, as a function of $\tau$, for
the same parameters used in the main panel. }
\label{fig:rvstfixgamma}
\end{figure}

\par We first focus on the case of perfect control ($\varepsilon=0$).
Representative results on the dependence of the improvement ratio $r$
upon the (minimum) gating interval $\tau$ are depicted in Fig.
\ref{fig:rvstfixgamma}, where the above $W_{\pi/4}^{(1,2)}$ gate is
considered for a fixed realization of the dipolar-bath Hamiltonian
$H_{B}$ \cite{Realizations}.  Weak- to strong- coupling decoherence
regimes are then probed by varying the ratio $A/\Gamma$. All curves
demonstrate the expected improvement in the limit of sufficiently
small $\tau$. Different features are worth highlighting: first, as
$\tau\rightarrow0$, the data are consistent with the quadratic scaling
with $1/\tau$ expected from Eq. (\ref{ratio}) (for $A=1$, for
instance, a fit of $\log(r)$ vs. $\log(\tau)$ yields
$1.9991\pm0.0002$). Emergence of such a scaling may be taken as a good
indication that asymptotic conditions are well obeyed by the
implementation. Second, the minimum gate interval
$\tau=\tau^{\ast}(A;\Gamma)$, at which the improvement region for DQEC
($r>1$) is entered as $\tau$ is decreased, depends on the system-bath
coupling strength -- a larger $A$ (stronger decoherence) requires
shorter values of $\tau^{\ast}$ to be accessed in order for DQEC to be
effective, as intuitively expected.  Note that, qualitatively, a
dependence of $r$ upon both $H_{SB}$ and $H_{B}$ is expected on the
basis of the uncorrected error associated with a DCG,
Eq. (\ref{eq:undecerr}). Only in the limit where $\Gamma\rightarrow0$
(a so-called `non-dynamical' bath \cite{Wen01,Wen03}) does the
first-order improvement regime simply relate to the coupling strength
$\norm{H_{SB}}$, which is fixed through $A$.

\begin{figure}[t]

\begin{centering}
\includegraphics[width=2.8in]{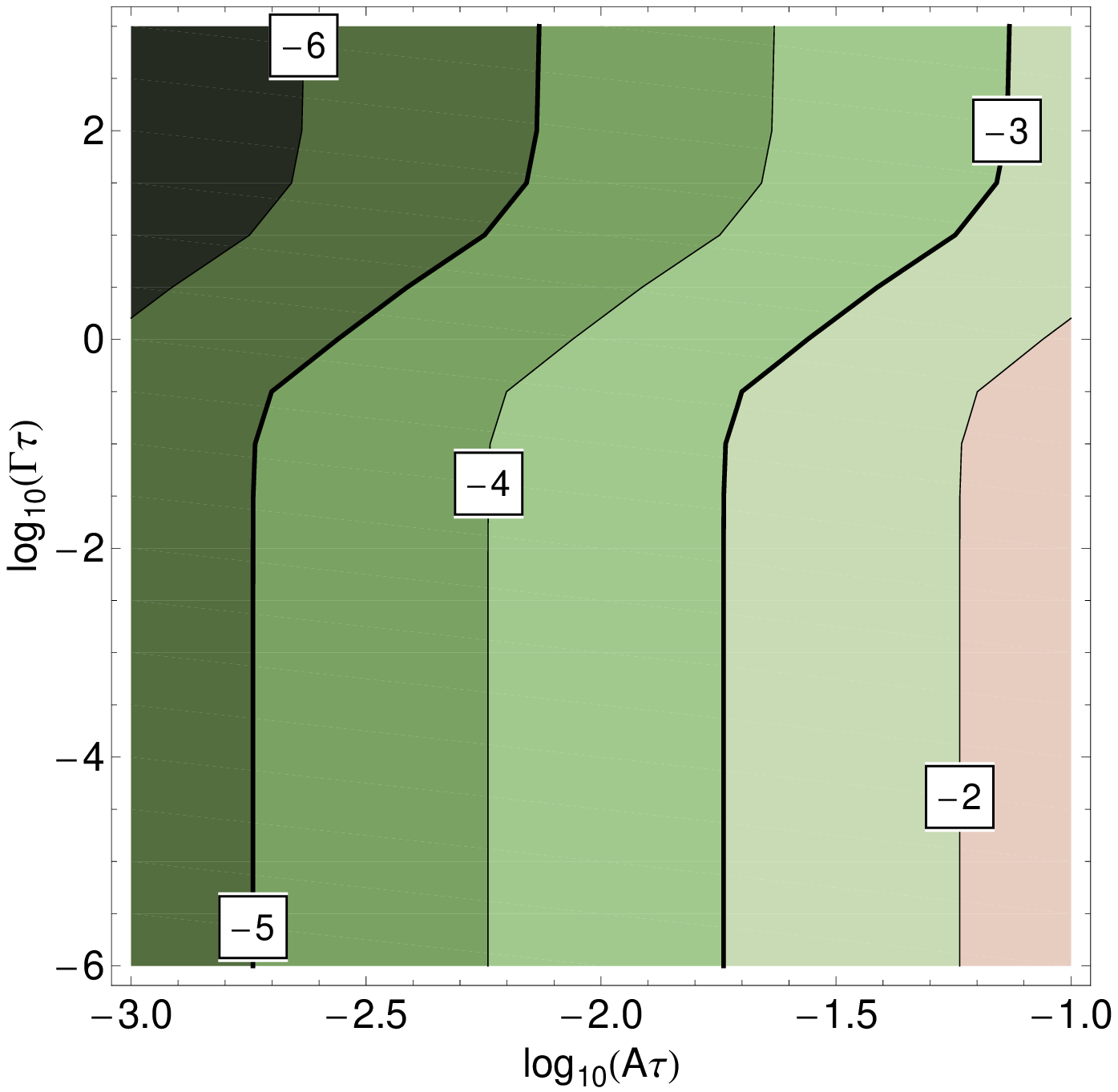}\\
\includegraphics[width=2.8in]{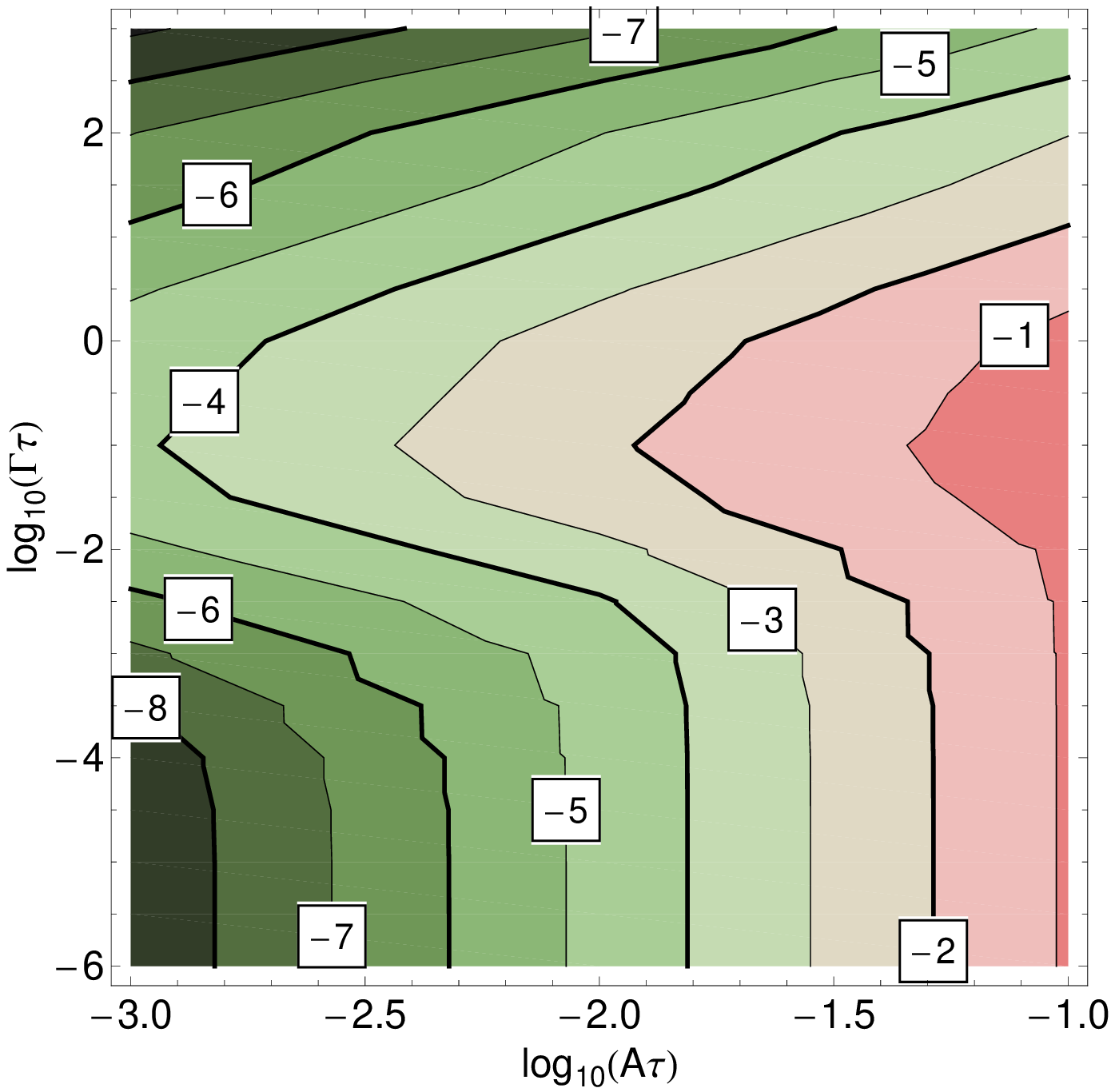}
\par\end{centering}

\caption{(color online) Iso-fidelity contour plots for the
implementation of the single-qubit target gate $R_{\pi/4}^{(1)}$ for
fixed (minimum) gating time $\tau=1$. Top: Primitive uncorrected
implementation.  Bottom: DCG implementation. In both cases, the
integers labeling each curve give (the logarithm of) the corresponding
infidelity error, with darker colors corresponding to higher fidelity
values. }
\label{fig:contour}
\end{figure}

\par Exploring the interplay between $H_{SB}$ and $H_{B}$ as they are
simultaneously varied for fixed (finite) values of the gating interval
$\tau$ reveals additional interesting structure, as shown in Fig.
\ref{fig:contour}. In particular, the loss of fidelity of a primitive
single-qubit gate (top panel) is contrasted to the loss of fidelity of
the corresponding DCG version (bottom panel) over a range of
intra-bath and system-bath couplings, $\Gamma$ and $A$,
respectively. Neither the uncorrected nor the DCG gate implementations
depend in a similar manner upon the two parameters, the asymmetry in
behavior being `amplified' by the DCG. As a first interesting feature,
only a mild dependence upon $\Gamma$ is observed for primitive gates
implementations, the corresponding fidelity \emph{increasing} for very
strong intra-bath dynamics. This effect can be partly understood if,
as mentioned in Sec. \ref{sub:dcgconstruction}, $H_{B}$ is included in
the definition of the toggling frame: A very strong $H_{B}$ will then
induce fast oscillations in $H_{e}(t)$, resulting in smaller EPGs
{[}{cf.}  Eq. (\ref{eq:Het}){]}. Secondly, and more interestingly, a
\emph{non-monotonic} behavior emerges for DCG implementations as the
intra-bath coupling $\Gamma$ is varied.

\begin{figure}[tbh]

\begin{centering}
\includegraphics[width=2.8in]{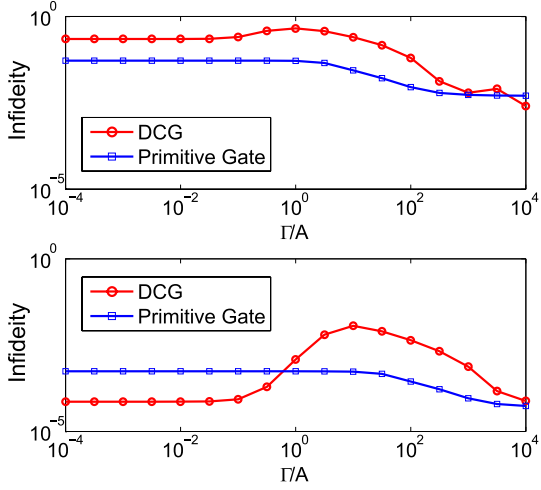}
\par\end{centering}

\caption{(color online) Infidelity, $1-f$, as a function of
intra-bath/system-bath coupling scales for implementing the
single-qubit gate $R_{\pi/4}^{(1)}$, at fixed $A=1$. Primitive vs DCG
implementations are contrasted for two different gating times
$\tau=0.1$ (top) and $\tau=0.01$ (bottom). }
\label{fig:probebath}
\end{figure}

The dependence of fidelities for DCG and primitive gates on $H_{B}$
through $\Gamma$ is further highlighted in Fig. \ref{fig:probebath},
where two representative values of $\tau$ are examined. The
resonance-like behavior in the DCG patterns is evident, even when no
significant improvement is expected. DQEC strategies such as DD or
DCGs thus may probe the strength of the internal bath dynamics
$H_{B}$, making them potentially attractive as a \emph{diagnostic
tool} for complex non-Markovian open quantum systems.

Besides the error effects associated with $H_{SB}$ and $H_{B}$, it is
instructive to explore the impact of classical control errors in our
setting, for example focusing on overrotation errors in the
single-qubit gate $R_{\pi/4}^{(1)}$. As further discussed in
Sec. \ref{sub:Hybridcomp} below, two different models for control
imperfections may be physically relevant: (i) The nominal control
Hamiltonian may be modified through the addition of a `deviation'
Hamiltonian with a fixed strength,
$H'_{\text{gate}}(t)=H_{\text{gate}}(t)+\varepsilon H_{\text{dev}}$,
where $H_{\text{dev}}$ is a Hermitian operator with units of energy
and $\varepsilon$ a small number (recall that $H_{\text{gate}}(t) =
H_{\text{ctrl}}(t)$ as long as $H_S=0$).  This results in a
\emph{scaled systematic EPG}, that can be tolerated (to first order)
DCG constructions, similar to the errors due to the bath.  (ii) The
nominal control Hamiltonian may be systematically scaled by a factor:
\begin{equation}
H'_{\text{gate}}(t)=(1+\varepsilon)H_{\text{gate}}(t).
\label{eq:syseps}
\end{equation}
\noindent
This results in a \emph{fixed systematic EPG}, in the sense that even
in the absence of the bath, a control segment of duration $\tau$
intended to produce $\exp(-i\theta C)$ will instead implement
$\exp[-i\theta(1+\varepsilon)C]$, regardless of $\tau$.  In the
simulations, model (ii) is implemented, given that there is no \emph{a
priori} indication that such errors should be tolerated by DCGs,
unlike errors of type (i).

\par
\begin{figure}[tb]
\begin{centering}
\includegraphics[width=3in]{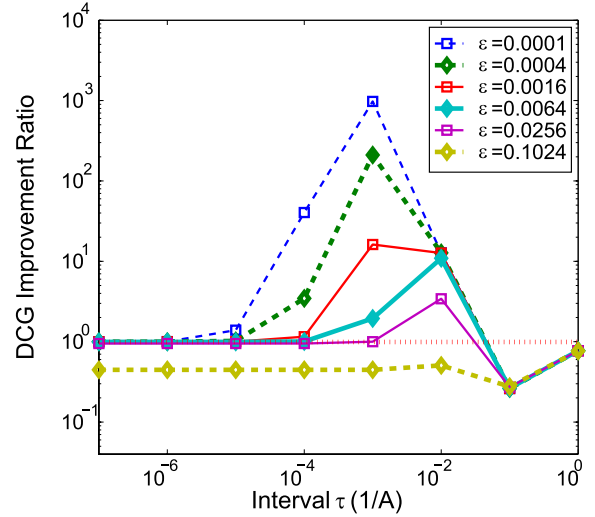}
\end{centering}
\caption{(color online) Effect of systematic over-rotation errors in
the DCG improvement ratio for $Q=R_{\pi/4}^{(1)}$ as a function of
gating interval $\tau$. In both cases, the control error strength is
parametrized by $\varepsilon\in[0,1]$, and we have set $A=1$,
$\Gamma=0$.}
\label{fig:contrerr}
\end{figure}

 Illustrative results for the DCG improvement ratio $r$ obtained
for a non-dynamical bath ($\Gamma=0$) at fixed $A=1$ are summarized in
Fig. \ref{fig:contrerr}, as a function of the gating interval and for
different values of the control error strength.
Overall, although the observed performance remains fairly stable for
error strengths up to a few percents, no intrinsic fault-tolerance is
observed for arbitrary strengths and/or rotation angles. In
particular, no improvement is achieved if uncorrected control errors
dominate over the decoherence errors (due to the spin-bath in this
case) that DGCs remove, as anticipated in \cite{khodjasteh-2008}.  For
small values of $\varepsilon$ (roughly below $1\%$), there is a set of
values of $\tau$ for which the decoherence error is indeed the leading
source of gate error, and for this region $r$ consistently exceeds
1. For the remaining values of $\tau$, and/or too large values of
$\varepsilon$, no improvement is observed: either $\tau$ is too large
for the asymptotic DCG regime to set in {(}cf. the interval
$[\tau^{*},\infty]$ in Fig. \ref{fig:rvstfixgamma}, as discussed
earlier{)}, or $\tau$ is too small and the decoherence errors become
negligible compared to the uncorrected control errors.  In the latter
case, no benefit arises from further reducing $\tau$, and performance
saturates at an $\varepsilon$-dependent limiting value. Qualitatively
similar results may be obtained for random control errors.

\section{Extension to faulty controls and systems with drift}
\label{sub:extenapp}

The explicit constructions of Sec. \ref{sec:dcgbeyond} for DCGs rely
on a straightforward relationship between the intended gate $Q$ and
the EPG associated with it. This is enabled by assumptions
\texttt{(a1)} and \texttt{(a2)} in Sec. \ref{sub:errassu}: we have
focused on $H_{SB}$ (causing decoherence) as the sole source of
(decoherence) errors, and the gating mechanism neither employed nor
was corrupted by drift terms in $H_{S}$.  In reality, as mentioned,
classical errors due to control imperfections and additional quantum
errors due to non-trivial unitary evolution are likely to be present
to a lesser or greater extent.  As also mentioned, robust control
methods have been devised for separately tackling these errors, whose
impact may be deleterious even if isolation from the environment is
good hence decoherence is not a primary concern.  In a way, composite
pulse techniques for systematic control errors
\cite{Jones-Composite,Brown2004,JonesComp,KhanejaUR} may themselves be
considered as a classical dynamical error correction strategy: they
are perturbative in nature; they assume no quantitative knowledge of
the value of errors and are constructed from building blocks of
primitive pulses. Likewise, strongly modulating and gradient-ascent
pulse engineering techniques in NMR
\cite{Fortunato-Control,Boulant03,KhanejaGRAPE} share with DQEC the
basic idea of suppressing unwanted (but known) coupled-spin dynamics
through coherent time-averaging, numerical search being explicitly
invoked to synthesize an optimal modulation.

The above considerations prompt the need for exploring DQEC under more
relaxed control assumptions than \texttt{(a1)} and \texttt{(a2)}.  In
particular, a natural question is whether it is possible to produce
{\em hybrid} dynamical error correction strategies to simultaneously
counter different types of classical and quantum errors, over which a
different degree of knowdlege may be available to begin with.  While
this question is too broad in scope to be answered in full here, and
might be best treated in conjunction with numerical optimization
approaches, we nonetheless illustrate the flexibility of analytic DQEC
methods by providing three specific examples of DCG DCG constructions
beyond \texttt{(a1)} and \texttt{(a2)}.

\subsection{Towards hybrid composite-pulse/ \\
DCG constructions}
\label{sub:Hybridcomp}

A large class of systematic control errors may be naturally modeled at
the Hamiltonian level.  That is, recalling Eq. (\ref{naive}), we may
assume that the nominal gating Hamiltonian $H_{\text{gate}}(t)$ during
the interval $[t_{1},t_{2}]$ is modified as
\begin{equation}
H'_{\text{gate}}(t)=H_{\text{gate}}(t)+H_{S,g}+ \varepsilon
H_{\text{\text{dev}}}(t),
\label{sys1}
\end{equation}
\noindent
where $H_{\text{dev}}(t)$ is (typically gate-dependent) Hamiltonian
whose functional form is known, and $\varepsilon$ is (possibly
unknown) real number such that \[
\norm{U_{\text{gate}}(t_{2},t_{1})-T_{+}\Big[\!\exp
\Big(\!\!\!-i\!\int_{t_{1}}^{t_{2}}\!\!\!
H_{\text{ctrl}}(s)ds\Big)\Big]}\ll 1.\]
\noindent
The systematic nature of the error is implicit in the existence of a
functional dependence between $H'_{\text{gate}}(t)$ and
$H_{\text{dev}}(t)$ as a mapping between time-dependent functions. For
physically relevant error models, this mapping is often specified in
terms of simple relationships between the control and the error
Hamiltonian -- {\em e.g.}, for a single qubit, application of an $X$
gating Hamiltonian may induce a `parallel' (in-axis) deviation
Hamiltonian also proportional to $X$, or a `perpendicular' (off-axis)
deviation Hamiltonian proportional to $Y,Z$ (see also
\cite{Wen03,Santos2008}).

\par More generally, a relatively simple description of systematic
control errors may be established by directly incorporating and
examining their effects into the error action operator.  Thus for each
target gate $Q$ ideally implemented via $H_{\text{gate}}(t)$ but
actually implemented via $H'_{\text{gate}}(t)$ in the presence of the
bath, to first order in the gate duration we can expand the associated
total error action $\Phi$ as:
\begin{equation}
\Phi=\Phi_{SB}+\Phi_{\text{ctrl}},
\label{sys2}
\end{equation}
where for simplicity we have assumed that no contribution arises from
$H_S$ (or that the latter commutes with $H_{SB}$ and
$\Phi_{\text{ctrl}}$ accounts for systematic control errors. The
assumptions \texttt{(a1)}-\texttt{(a2)} allowed us to let
$\Phi_{\text{ctrl}}=0$, and provided indirect control over $\Phi_{SB}$
through the gating interval $\tau$: For primitive gates in the absence
of non-commutative drift terms, $\Phi_{SB}$ is linearly proportional
to $\tau$. Such a dependence of $\Phi_{SB}$ on $\tau$ could then be
explicitly utilized in the constructions of Sec. \ref{sub:diffequal}
for matching the errors in $QQ'$ and $Q_{1/2}$.

\par Fixed systematic errors {[}cf. Eq. (\ref{eq:syseps}){]} refer to
$\Phi_{\text{ctrl}}$ terms in Eq. (\ref{sys2}) that do {\em not} scale
with $\tau$, as opposed to scaled systematic errors that scale
linearly with $\tau$. Hamiltonian-control errors as described by
Eq. (\ref{sys1}) always lead to scaled systematic errors.  Since the
latter can be formally absorbed into $\Phi_{SB}$, in the absence of
additional (drift) errors, the constructions of
Sec. \ref{sub:diffequal} may still be used to effectively reduce the
effect of systematic control errors in the same way the effects of the
bath are corrected.  That is, arbitrary DCGs are intrinsically robust
against scaled systematic errors.  The EDD construction for {\sc noop}
on the other hand (as a special DCG), is tolerant of {\em all}, scaled
and not, systematic control errors (as long as they are small), for
they as can be absorbed into generator gate errors $\Phi_{F_{_{i}}}$
\cite{Viola2003Euler}.  The numerical results of Sec. \ref{sub:num}
confirm instead the expectation that generic DCG gates need not be
robust under fixed systematic errors.

It is still plausible that one may use a combination of composite
pulses and DCGs to actively stabilize quantum gates against both
decoherence and systematic errors. The starting point is a composite
pulse construction that in the absence of the environment can generate
a desired gate $Q$ robustly with respect to systematic errors within a
specific model.  Let us assume that it is possible to modify the
control inputs for this construction to obtain the gates $Q_{1/2}$ and
$Q^{'}$ defined in Sec. \ref{sec:dcgbeyond}, such that $Q_{1/2}$ and
$Q'$ constructed in this way are also composite pulse robust against
the same control errors.  Then clearly the DCG construction that
employs $Q$, $Q'$, $Q_{1/2}$, along with the generator gates $F_{i}$,
will correct (to first order) the systematic errors and the errors due
to the bath. This, however, necessarily involves longer and more
complex sequences of primitive control Hamiltonians.

\subsection{Qubits in rotating frames}

An important class of systems with drift are qubits which are defined
in terms of two (non-degenerate) internal energy levels and are
addressed using resonance techniques. This is, for instance, the
typical setting in both liquid- and solid- state NMR QIP devices
\cite{Cory-Overview1}.  In this case, the internal open-system
Hamiltonian $H_{\text{int}}$ includes `chemical-shift' terms of the
form $H_{S}=\sum_{i} \Omega_i Z_{i}$, which are employed for
single-qubit manipulations by moving to a frame which rotates with the
applied carrier frequency $\Omega_c$ and by tuning on-resonance with
the target spin(s). The rotating frame transformation has the effect
of introducing an explicit time-dependence in system (hence
system-bath) operators, according to
$$ {\cal O} \mapsto\tilde{\cal O}(t,t_0)=\exp(-it \Omega_c \sum_i
Z_i)\, {\cal O}\, \exp(it \Omega_c \sum_i Z_i ), $$
\noindent
where we have assumed that $t_0=0$.
Thus, even neglecting contributions to $\tilde{H}_S$ from
off-resonance qubits, both the interaction with the environment and
therefore the error Hamiltonian are transformed, in general, into
carrier-modulated Hamiltonians $\tilde{H}_{SB}(t,0)$ and
$\tilde{H}_{e}(t,0)$, respectively.  This causes most of the
constructions in the current paper (including EDD) to be not
immediately applicable to compensate the rotating-frame error
associated with a gate, because this error will generally depend on
the time at which the gate is applied. Still, thanks to the known
periodicity of the time dependence of the error actions, it is
possible to {\em synchronize} the duration of the applied primitive
gates with the rotating frame period $2\pi/\Omega$, in such a way that
the relationships between the errors assumed in EDD and DCG
constructions are {\em stroboscopically} preserved.

Furthermore, in practice the frequency $\Omega_c$ may be much higher
than typical inverse gate durations.  This results in rapidly
oscillating terms in the rotating franes that are effectively canceled
[when Eq. (\ref{eq:errortoggle}) is used with $\tilde{H}_{e}(t)$]. If
$H_{e}$ consists of arbitrary single-qubit error terms
{[}Eq. (\ref{eq:Hsbsingle}){]}, then in the limit of high carrier
frequencies, the effective error model for the open-system dynamics in
the rotating frame approaches the pure dephasing error model, since
any terms proportional to $X^{(i)}$ and $Y^{(i)}$ in the error action
will be averaged to zero, whereas any terms proportional to $Z^{(i)}$
will be time-independent. This has two implications: On the one hand,
single-qubit drift terms in $H_S$ are of no concern for DQEC in the
important case where the physical error model is dephasing to begin
with; on the other hand, if arbitrary decoherence is the appropriate
model in the physical frame, DCG constructions for pure dephasing are
directly applicable in the rotating frame in the limit of large
$\Omega_c$.

\subsection{Always-on qubit-qubit interactions}

Aside from single-qubit drift terms as discussed above, another common
type of drift contributions may arise from always-on two-qubit
couplings in $H_S$.  In NMR QIP, for instance, the latter can either
be `$J$-coupling' Ising-like Hamiltonians for weakly interacting spin
systems, or Heisenberg-like Hamiltonians in the strong-coupling
regimes.
Within our present approach, the main complication that a
non-switchable component $H_{S,g}$ in the gating Hamiltonian for $Q$
introduces towards constructing a DCG version is in establishing the
existence of the required $Q_{1/2}$ and $Q'$ gates, since $H_{S,g}$
effectively imposes a limit on the range of achievable control inputs.
Specifically, the constructions of Sec. \ref{sec:dcgbeyond} depend on
generating propagators $U'_{\text{gate}}(t,0)$ and
$U''_{\text{gate}}(t,0)$, and it is not immediately clear whether
universal control over the system $S$ suffices for realizing such
propagators (as a one-parameter family of time-dependent operators as
opposed to just end-point unitaries), even in an approximate sense.
Nonetheless, we see no fundamental obstacle to treating such cases
through strong modulation of the system's dynamics and numerical
optimization.  To further support our optimism, we show here how
analytical DCG constructions remain feasible upon careful
consideration of the relevant constraints on a case-by-case basis.

\par Consider a system of $n$ Heisenberg-coupled qubits in a
nearest-neighbor configuration:
\begin{equation}
H_{S}=\lambda\sum_{i=1}^{n-1}\mathbf{S}^{(i)}\cdot\mathbf{S}^{(i+1)},
\label{eq:alwaysHeis}
\end{equation}
\noindent
where $\lambda \in {\mathbb R}$ is a fixed coupling strength.  Given
$H_{S}$ in Eq. (\ref{eq:alwaysHeis}), universal control over $S$ can
be gained by assuming access to a switchable control Hamiltonians
of the following form:
\begin{equation}
H_{\text{ctrl}}(t)=\sum_{i}[h_{X}^{(i)}(t)X^{(i)}+h_{Y}^{(i)}(t)Y^{(i)}],
\label{eq:singcont}
\end{equation}
\noindent
where as usual the control inputs $h_{\alpha}^{(i)}(t)$ are subject to
the power and bandwidth constraints \texttt{(c1)}-\texttt{(c2)}.
We assume the bath to interact with $S$ according to the arbitrary
linear decoherence model: $H_{SB}\in\Omega^{\{1\}}_e$. Our goal is to
provide a (first-order) DCG construction for the following gates:
$\exp[-i\theta S_{\alpha}^{(k)}]$, where $\alpha=X,Y$, and
$\exp[-i\theta\mathbf{S}^{(k)}\cdot\mathbf{S}^{(k+1)}]$, such that the
resulting EPG scales with $\tau_{\min}^{2}$. Our methodology here will
be somewhat different from Sec. \ref{sec:dcgbeyond}, in the sense that
primitive gates are not directly used as the basis of DCG
constructions.

\par We first make the following general observation: Suppose that the
gating propagator and the error Hamiltonian for a control block in the
interval $[0,T]$ can be respectively written as
$U_{\text{gate}}(0,t)=U_{g,1}(t)U_{g,2}(t)$ and
$H_{e}=H_{e,1}+H_{e,2}$, in such a way that
\begin{eqnarray}
&&[U_{g,1}(t),U_{g,2}(t')]  =  0,\label{eq:ug1ug2}\\{}
&&[U_{g,1}(t),H_{e,2}] =  [U_{g,2}(t),H_{e,1}]  =0.
\label{eq:ughe}
\end{eqnarray}
\noindent
Using the above equations and the tools of Sec. \ref{sec:tools}, we
can write the error associated with this block up to the first order
as $\Phi^{[1]} = \Phi_{1}^{[1]}+\Phi_{2}^{[1]}$, where
\begin{eqnarray}
\Phi_{1}^{[1]} & = &
\int_{0}^{T}U_{g,1}(t)^{\dagger}H_{e,1}U_{g,1}(t)dt,
\label{divided1} \\ \Phi_{2}^{[1]}& = &
\int_{0}^{T}U_{g,2}(t)^{\dagger}H_{e,2}U_{g,2}(t)dt.
\label{divided2}
\end{eqnarray}

\par As it turns out, generating a two-qubit gate on a target qubit
pairs, $\exp[-i\theta\mathbf{S}^{(k)}\cdot\mathbf{S}^{(k+1)}]$, is the
most non-trivial step.  In this case, we clearly have $H_{S,g}=
\mathbf{S}^{(k)}\cdot\mathbf{S}^{(k+1)}$ and $H_S-H_{S,g}= H_{S,e}$.
Basically, our strategy is to first devise a scheme for
dynamically correcting all error terms affecting any qubit other than
$k$ and $k+1$, and next to superpose a scheme that remove any
additional error term while steering the target qubits according to
$H_{S,g}$.  For the first step, consider the DD group
$\mathcal{G}_{\text{NN}}=\mathbb{Z}_{2}\otimes\mathbb{Z}_{2}
\otimes\mathbb{Z}_{2}\otimes\mathbb{Z}_{2}$, designated to decouple
the subspace of bilinear qubit terms
\[
\Omega_{k}^{\{2\}}=\{S_{\alpha}^{(i)}S_{\beta}^{(i+1)}\otimes
B_{\alpha}^{(i)}\},\]
\noindent
where $i=1,\cdots,k-1,k+1,\cdots,n-1$,
$B_{\alpha}^{(i)}\in\mathcal{B}(H_{B})$ and $S_{\beta}^{(i+1)}$ may
include the identity operator $I^{(i+1)}$.  $\mathcal{G}_{\text{NN}}$
has 16 elements, and 4 generators which we choose to represent in
${\cal H}_S$ via
\begin{equation}
\{F_{1,j}\}=\{X^{\text{odd}}X^{\text{ev}},X^{\text{odd}}Y^{\text{ev}},
\;\;\; Y^{\text{odd}}X^{\text{ev}},Y^{\text{odd}}Y^{\text{ev}}\},
\label{eq:F1j}
\end{equation}
\noindent
where
\[ S_{\alpha}^{\text{odd}}=\prod_{j\in
E_{k}}S_{\alpha}^{(j)},\;\; S_{\alpha}^{\text{ev}}=\prod_{j\in
O_{k}}S_{\alpha}^{(j)},\]
\begin{eqnarray*}
E_{k} & = &
\{j\,|\,j\text{ even};0<j\le n;j\ne k,k+1\},\\
O_{k} & = & \{j\,|\,j\text{odd};0<j\le n;j\ne k,k+1\}.
\end{eqnarray*}
\noindent
Consider the EDD sequence based on $\{F_{i,1}\}$, and let $H_{g,1}(t)$
be the gating Hamiltonian associated with it over the interval
$[0,T_{\text{two}}]=[0,16\times 4\tau]$, assuming as usual that each
generator is implemented over an interval of length $\tau$.  Note that
$H_{g,1}(t)$ can be generated through appropriate combinations of the
switchable inputs in Eq. (\ref{eq:singcont}).

\par In the second step, we construct the entangling gate $\exp[-64 i
\tau\lambda\mathbf{S}^{(k)}\cdot\mathbf{S}^{(k+1)}]$ by applying
another control Hamiltonian $H_{g,2}(t)$ in parallel with
$H_{g,1}(t)$.  $H_{g,2}(t)$ implements a sequence very similar to the
EDD sequence for the group
$\mathcal{G}_{\text{LD}}=\mathbb{Z}_{2}\otimes\mathbb{Z}_{2}$
designated to cancel errors in the subspace \[
\Omega_{LD,k}=\{S_{\alpha}^{(i)}\otimes
B_{\alpha}^{(i)}\,|\,i=k,k+1\}.\] However, while the original EDD
sequence would employ the generators $X^{(k)}X^{(k+1)}$ and
$Y^{(k)}Y^{(k+1)}$ (or a similar combination), we now replace the two
original bilinear generators with
\begin{eqnarray*}
F_{1,2} & = & X^{(k)}X^{(k+1)}
\exp[-16i\tau\lambda\mathbf{S}^{(k)}\cdot\mathbf{S}^{(k+1)}],\\
F_{2,2} & = &
Y^{(k)}Y^{(k+1)}\exp[-16i\tau\lambda\mathbf{S}^{(k)}\cdot\mathbf{S}^{(k+1)}].
\end{eqnarray*}
\noindent
The modified EDD sequence can be implemented through the available
control Hamiltonians via a collective gating Hamiltonian of the form
\[
H_{g,2}(t)=H_{l}^{(k)}, \;\;\text{ for }t\in[8(l-1)\tau,8l\tau],\] with
\begin{equation}
H_{l}^{(k)}=\begin{cases}
\substack{\frac{\pi}{16\tau}[X^{(k)}+X^{(k+1)}] + H_{S,g} }
,\, & l=1,3,6,8,\\
\substack{\frac{\pi}{16\tau}[Y^{(k)}+Y^{(k+1)}] + H_{S,g} }
,\, & l=2,4,5,7.
\end{cases}
\label{eq:kig2}
\end{equation}
\noindent
Notice that the system gating Hamiltonian, explicitly included in
$H_{g,2}(t)$, commutes with the single qubit terms acting on $k$ and
$k+1$ in Eq. (\ref{eq:kig2}). Let
\begin{eqnarray*}
U_{g,l}(t) & = &
T_{+}\Big[\exp\Big(-i\int_{0}^{64\tau}\!H_{\text{g},l}(s)ds\Big)\Big],\\
H_{e,1} & = & \lambda\sum_{\substack{i=1\\ i\ne k}
}^{n}\mathbf{S}^{(i)} \cdot\mathbf{S}^{(i+1)} +
\sum_{\alpha}\!\sum_{\substack{i=1\\ i\ne k,k+1}
}^{n}\!\! S_{\alpha}^{(i)}\otimes B_{\alpha}^{(i)}+H_{B},\\
H_{e,2} & =& \sum_{\alpha} \sum_{i=k}^{k+1}S_{\alpha}^{(i)}\otimes
B_{\alpha}^{(i)}.
\end{eqnarray*}
\noindent
We may thus readily verify
Eqs. (\ref{eq:ug1ug2})\nobreakdash-(\ref{eq:ughe}).  Also notice that
the total gating Hamiltonian commutes with
$\mathbf{S}^{(k)}\cdot\mathbf{S}^{(k+1)}$.  Thus, the overall action
of the control block defined via $H_{g,1}(t)+H_{g,2}(t)$ is given by
$U_{g,1}(64\tau)U_{g,2}(64\tau)=
\exp[-64i\tau\lambda\mathbf{S}^{(k)}\cdot\mathbf{S}^{(k+1)}]$, and we
can derive the error associated with the whole sequence (of length
$T_{\text{two}} =64\tau$) by invoking Eqs. (\ref{divided1})-
(\ref{divided2}):
\begin{eqnarray*}
\Phi_{1}^{[1]} & = &
\int_{0}^{T}\!U_{g,1}(t)^{\dagger}H_{e,1}U_{g,1}(t)dt=\sum_{l}
\Phi_{1,l}^{[1]},\\
\Phi_{2}^{[1]} & = &
\int_{0}^{T}\!U_{g,2}(t)^{\dagger}H_{e,2}U_{g,2}(t)dt=\sum_{l}
\Phi_{2,l}^{[1]},
\end{eqnarray*}
\noindent
where $\Phi_{1,l}^{[1]}$ and $\Phi_{2,l}^{[1]}$ respectively denote the
errors due to $H_{e,1}$ and $H_{e,2}$, associated with interval
$[(l-1)\tau,l\tau]$.  By (EDD) construction
\[ \Phi_{1}^{[1]}=\sum_{j=1}^{4}
\Pi_{\mathcal{\mathcal{G}_{\text{NN}}}}(\Phi_{F_{1,j}}).\]
\noindent
One can verify directly {[}similar to Sec. \ref{sub:contham}{]} that
using $H_{g,1}(t)$ results in an error $\Phi_{F_{1,j}}$ associated
with each generator $F_{1,j}$ that belongs to $\Omega_{k}^{\{2\}}$,
and thus we have $\Phi_{1}^{[1]} =0\text{ mod}B$. Similarly, one may
also verify that the error $\Phi_{2,l}^{[1]}$ belongs to
$\Omega_{\text{LD},k}$.  We may then use Eq. (\ref{eq:PhiA1}) and the
fact that the commutator between
$\lambda\mathbf{S}^{(k)}\cdot\mathbf{S}^{(k+1)}$ and
$\sum_{i=k}^{k+1}S_{\alpha}^{(i)}\otimes B_{\alpha}^{(i)}$ can be
ignored up to $O[\tau^{2}\max(\Vert B_{\alpha}^{i)}\Vert)\lambda]$ to
write \[
\Phi_{2}^{[1]}=\sum_{j=1}^{2}\Pi_{\mathcal{G}_{\text{LD}}}
(\Phi_{F_{2,j}}),\]
\noindent
therefore $\Phi_{2}^{[1]}=0\text{ mod}B$. The total error associated
with this sequence is thus 0 modulo pure bath terms and terms that
scale with $\tau^{2}$, as desired.

\par Implementing a DCG version of a single-qubit rotation $Q$ on a
selected target qubit $k$ is conceptually more straightforward.  In
this case, $H_S\equiv H_{S,e}$ thus $H_{e}=H_{SB}+H_{S}$, and the DCG
construction of Sec. \ref{sec:dcgbeyond} can be reproduced with ease
by letting the relevant error subspace cover \emph{all}
nearest-neighbor bi-linear interactions of the form \[
\Omega^{\{2\}}_e=\{S_{\alpha}^{(i)}S_{\beta}^{(i+1)}\otimes
B_{\alpha}^{(i)}\},\]
\noindent
where $i=1,\cdots,n-1$ and $B_{\alpha}^{(i)}\in\mathcal{B}(H_{B})$. The
space $\Omega^{\{2\}}_e$ can be decoupled by the group
$\mathcal{G}_{\text{NN}}$ with a representation similar to
Eq. (\ref{eq:F1j}), except that now the sets $E_{k}$ and $O_{k}$ are
chosen as
\begin{eqnarray*}
E_{k} & = & \{j\,|\,j\text{ even};0<j\le n\},\\
O_{k} & = & \{j\,|\, j\text{ odd};0<j\le n\}.
\end{eqnarray*}
\noindent
Similarly, we may generate $Q$, $Q'$, and $Q_{1/2}$ by simply
switching the appropriate single-qubit Hamiltonian on the $k$-th
qubit. One may verify that the error associated with each control
segment belongs to $\Omega^{\{2\}}_e$.  The final DCG construction
will thus contain $64$ segments for EDD and $32$ segments for
$Q_{1/2}$ and $QQ'$, for a total duration of
$T_{\text{single}}=96\tau$.

\par It is worth noting that, in a way, the larger number of control
segments emerging from the above analysis as compared to driftless
scenarios may be simply regarded as an analytically generated
`digitized' pulse shape, with a basic time-step determined by
$\tau_{\text{min}}$.
Attention must be paid, however, to the fact that longer control
sequences might potentially invalidate
the (sufficient) convergence assumptions implicit in DCG
constructions.  This highlights the importance of finding an {\em
optimal} DCG construction for a given control scenario.  In this
sense, the constructions of this section should be considered as a
proof-of-concept as opposed to optimal results that might best be
tailored to specific control scenarios and will likely involve a
combination of analytical and numerically optimized control designs.

\section{Conclusion and outlook}
\label{theend}

\par We have provided the mathematical and control-theoretic framework
for DCGs as introduced in Ref. \onlinecite{khodjasteh-2008}, and
further explored their range of applicability.  Besides examining in
full detail the analytical construction of decoherence-protected gates
in the prototypical scenario of arbitrary single-qubit errors, a
physically motivated setting involving spin-bath decoherence has been
characterized through exact numerical simulation.  This has both
confirmed the expected performance of DCGs in a wide parameter regime
and yielded insight into the interplay of different open-system and
control features.  Additional challenges arising in the synthesis of
DGCs under realistic control assumptions involving drifts and control
errors have been laid out, and proof-of-concept solutions constructed
in specific representative situations.

\par Our present analysis naturally points to several directions for
further investigation.  First, our focus here has been on DCG
constructions that cancel errors only up to the first order in Magnus
expansion. In principle, it is possible to go beyond the first order
DCG constructions and produce even more efficient DQEC protocols
\cite{khodjasteh-2009}.  While this will unavoidably involve a higher
level of sophistication in the resulting control modulation and/or the
computational effort involved in numerical search processes, a clear
understanding of relevant complexity properties is needed.  Making
contact with the concept of `quantum control landspace' as recently
proposed for the generation of unitary transformations in closed
quantum systems \cite{RabitzLand} might prove especially insightful.

\par Second, we only considered a limited set of permissible control
Hamiltonians for DCG constructions. A general analysis of what we
referred to as the `gate permissibility problem' for a given open
quantum systems
is, as mentioned, still lacking. This generalization appears even more
interesting by observing that the set of permissible controls can be
directly related to the specific representation of the DD group used
in the DCG construction.  In fact, the apparent freedom in the choice
of DD representation independently motivates the
existence of DQEC proposals for the same tasks as DCGs that do not
rely upon DD explicitly. Numerical optimization is, in this sense, a compelling
option to consider for investigating more general DQEC constructions.

\par Lastly, different options exist for incorporating DQEC protocols
into actual QIP architectures.  While we have stressed the potential
usefulness of DCGs directly at the {\em physical} level, DCG
constructions could potentially be envisioned on top of a {\em
logical} encoding, for instance to reduce EPGs in silicon
double-quantum-dot logical qubits as considered in \cite{Levy09}.
Once an algorithm is specified, whether at the physical or logical
level, one may further consider whether to produce the DCG version of
each circuit gate, and obtain an error-suppressed algorithm
accordingly; or whether instead to dynamically correct the whole
algorithm viewed as a complex unitary transformation.
While a thorough understanding of such design issues is fundamental
for real-world applications, optimal solutions will ultimately depend
on various details and constraints in place and, as such, are most
effectively addressed in specific contexts. It is our hope that this
work will prompt further interest and investigation in DQEC approaches
from the QIP implementation community.

\begin{acknowledgments}

It is a pleasure to thank Daniel Lidar and Andrew Landahl for
insightful questions and valuable feedback. K.K. and L.V. gratefully
acknowledge support from the National Science Foundation under Grants
No. PHY-0555417 and PHY-0903727, the Department of Energy, Basic
Energy Sciences, under Contract No.  DE-AC02-07CH11358, and from
Constance and Walter Burke through their Special Projects Fund in
Quantum Information Science.

\end{acknowledgments}

\appendix
\section{Numerical methods}

\label{app}

The simulations described in Sec. \ref{sub:num} were performed (in
{\textsc{Mathematica}}) using direct exact exponentiation for
obtaining the propagators describing the system-plus-bath dynamics in
the presence of control. Arbitrary precision matrices turned out to be
necessary for the following reasons: (i) Small errors correspond to
toggling-frame error propagators that are close to identity, thus it
is essential that any numerical error be much smaller than the
physical effect being probed. (ii) Each matrix multiplication
effectively reduces the precision of the calculation by a fixed amount
which depends on the dimension of the matrices used. We were limited
to consider a relatively small system/bath and use matrices of very
high precision {[}60 digits or higher{]} throughout the computations.


\begin{thebibliography}{71}
\expandafter\ifx\csname natexlab\endcsname\relax\def\natexlab#1{#1}\fi
\expandafter\ifx\csname bibnamefont\endcsname\relax
  \def\bibnamefont#1{#1}\fi
\expandafter\ifx\csname bibfnamefont\endcsname\relax
  \def\bibfnamefont#1{#1}\fi
\expandafter\ifx\csname citenamefont\endcsname\relax
  \def\citenamefont#1{#1}\fi
\expandafter\ifx\csname url\endcsname\relax
  \def\url#1{\texttt{#1}}\fi
\expandafter\ifx\csname urlprefix\endcsname\relax\def\urlprefix{URL }\fi
\providecommand{\bibinfo}[2]{#2}
\providecommand{\eprint}[2][]{\url{#2}}

\bibitem[{\citenamefont{DiVincenzo}(2000)}]{DiVincenzo-Overview}
\bibinfo{author}{\bibfnamefont{D.~P.} \bibnamefont{DiVincenzo}},
  \bibinfo{journal}{Fortschr. Phys.} \textbf{\bibinfo{volume}{48}},
  \bibinfo{pages}{771} (\bibinfo{year}{2000}).

\bibitem[{\citenamefont{Nielsen and Chuang}(2000)}]{NielsenBook}
\bibinfo{author}{\bibfnamefont{M.~A.} \bibnamefont{Nielsen}} \bibnamefont{and}
  \bibinfo{author}{\bibfnamefont{I.~L.} \bibnamefont{Chuang}},
  \emph{\bibinfo{title}{Quantum Computation and Quantum Information}}
  (\bibinfo{publisher}{Cambridge University Press},
  \bibinfo{address}{{Cambridge, UK}}, \bibinfo{year}{2000}).

\bibitem[{\citenamefont{Kitaev}(1997)}]{Kitaev-Algorithms}
\bibinfo{author}{\bibfnamefont{A.~Y.} \bibnamefont{Kitaev}},
  \bibinfo{journal}{Russ. Math. Surv.} \textbf{\bibinfo{volume}{52}},
  \bibinfo{pages}{1191} (\bibinfo{year}{1997}).

\bibitem[{\citenamefont{Preskill}(1998)}]{PreskillRel}
\bibinfo{author}{\bibfnamefont{J.}~\bibnamefont{Preskill}},
  \bibinfo{journal}{Proc. R. Soc. London A} \textbf{\bibinfo{volume}{454}},
  \bibinfo{pages}{385} (\bibinfo{year}{1998}).

\bibitem[{\citenamefont{Knill et~al.}(1998)\citenamefont{Knill, Laflamme, and
  Zurek}}]{Knill-Resilient}
\bibinfo{author}{\bibfnamefont{E.}~\bibnamefont{Knill}},
  \bibinfo{author}{\bibfnamefont{R.}~\bibnamefont{Laflamme}}, \bibnamefont{and}
  \bibinfo{author}{\bibfnamefont{W.~H.} \bibnamefont{Zurek}},
  \bibinfo{journal}{Science} \textbf{\bibinfo{volume}{279}},
  \bibinfo{pages}{342} (\bibinfo{year}{1998}).

\bibitem[{\citenamefont{Steane}(2003)}]{Steane03}
\bibinfo{author}{\bibfnamefont{A.~M.} \bibnamefont{Steane}},
  \bibinfo{journal}{Phys. Rev. A} \textbf{\bibinfo{volume}{68}},
  \bibinfo{pages}{042322} (\bibinfo{year}{2003}).

\bibitem[{\citenamefont{Knill}(2005)}]{Knill-Noisy}
\bibinfo{author}{\bibfnamefont{E.}~\bibnamefont{Knill}},
  \bibinfo{journal}{Nature} \textbf{\bibinfo{volume}{434}}, \bibinfo{pages}{39}
  (\bibinfo{year}{2005}).

\bibitem[{\citenamefont{Haeberlen}(1976)}]{HaeberlenBook}
\bibinfo{author}{\bibfnamefont{U.}~\bibnamefont{Haeberlen}},
  \emph{\bibinfo{title}{High Resolution NMR in Solids: Selective Averaging}}
  (\bibinfo{publisher}{Academic Press}, \bibinfo{address}{New York},
  \bibinfo{year}{1976}).

\bibitem[{Cor()}]{Cory-Overview1}
\bibinfo{note}{D. G. Cory {\em et al.}, Fortschr. Phys. {\bf 48}, 875 (2000).}

\bibitem[{\citenamefont{Ryan et~al.}()\citenamefont{Ryan, Negrevergne,
  Laforest, Knill, and Laflamme}}]{Ryan08}
\bibinfo{author}{\bibfnamefont{C.}~\bibnamefont{Ryan}},
  \bibinfo{author}{\bibfnamefont{C.}~\bibnamefont{Negrevergne}},
  \bibinfo{author}{\bibfnamefont{M.}~\bibnamefont{Laforest}},
  \bibinfo{author}{\bibfnamefont{E.}~\bibnamefont{Knill}}, \bibnamefont{and}
  \bibinfo{author}{\bibfnamefont{R.}~\bibnamefont{Laflamme}},
  \bibinfo{note}{eprint arXiv:0803.1982}.

\bibitem[{\citenamefont{Viola and Lloyd}(1998)}]{Viola1998}
\bibinfo{author}{\bibfnamefont{L.}~\bibnamefont{Viola}} \bibnamefont{and}
  \bibinfo{author}{\bibfnamefont{S.}~\bibnamefont{Lloyd}},
  \bibinfo{journal}{Phys. Rev. A} \textbf{\bibinfo{volume}{58}},
  \bibinfo{pages}{2733} (\bibinfo{year}{1998}).

\bibitem[{\citenamefont{Viola et~al.}(1999{\natexlab{a}})\citenamefont{Viola,
  Knill, and Lloyd}}]{Viola1999Dec}
\bibinfo{author}{\bibfnamefont{L.}~\bibnamefont{Viola}},
  \bibinfo{author}{\bibfnamefont{E.}~\bibnamefont{Knill}}, \bibnamefont{and}
  \bibinfo{author}{\bibfnamefont{S.}~\bibnamefont{Lloyd}},
  \bibinfo{journal}{Phys. Rev. Lett.} \textbf{\bibinfo{volume}{82}},
  \bibinfo{pages}{2417} (\bibinfo{year}{1999}{\natexlab{a}}).

\bibitem[{\citenamefont{Viola and Knill}(2005)}]{Viola2005Random}
\bibinfo{author}{\bibfnamefont{L.}~\bibnamefont{Viola}} \bibnamefont{and}
  \bibinfo{author}{\bibfnamefont{E.}~\bibnamefont{Knill}},
  \bibinfo{journal}{Phys. Rev. Lett.} \textbf{\bibinfo{volume}{94}},
  \bibinfo{pages}{060502} (\bibinfo{year}{2005}).

\bibitem[{\citenamefont{Santos and Viola}(2006)}]{Santos2006}
\bibinfo{author}{\bibfnamefont{L.~F.} \bibnamefont{Santos}} \bibnamefont{and}
  \bibinfo{author}{\bibfnamefont{L.}~\bibnamefont{Viola}},
  \bibinfo{journal}{Phys. Rev. Lett.} \textbf{\bibinfo{volume}{97}},
  \bibinfo{pages}{150501} (\bibinfo{year}{2006}).

\bibitem[{CDD()}]{CDD}
\bibinfo{note}{K. Khodjasteh and D. A. Lidar, Phys. Rev. Lett. {\bf 95}, 180501
  (2005); Phys. Rev. A {\bf 75}, 062310, (2007).}

\bibitem[{UDD()}]{UDD}
\bibinfo{note}{G. S. Uhrig, Phys. Rev. Lett. {\bf 98}, 100504 (2007);
{\em ibid.} {\bf 102}, 120502 (2009).}

\bibitem[{\citenamefont{Berglund}()}]{Berglund2000}
\bibinfo{author}{\bibfnamefont{A.~J.} \bibnamefont{Berglund}},
  \bibinfo{note}{eprint arXiv:quant-ph/0010001}.

\bibitem[{\citenamefont{Fraval et~al.}(2005)\citenamefont{Fraval, Sellars, and
  Longdell}}]{Fraval}
\bibinfo{author}{\bibfnamefont{E.}~\bibnamefont{Fraval}},
  \bibinfo{author}{\bibfnamefont{M.~J.} \bibnamefont{Sellars}},
  \bibnamefont{and} \bibinfo{author}{\bibfnamefont{J.~J.}
  \bibnamefont{Longdell}}, \bibinfo{journal}{Phys. Rev. Lett.}
  \textbf{\bibinfo{volume}{95}}, \bibinfo{pages}{030506}
  (\bibinfo{year}{2005}).

\bibitem[{\citenamefont{Morton et~al.}(2006)\citenamefont{Morton, Tyryshkin,
  Ardavan, Benjamin, Porfyrakis, Lyon, and Briggs}}]{Morton:2006}
\bibinfo{author}{\bibfnamefont{J.}~\bibnamefont{Morton}},
  \bibinfo{author}{\bibfnamefont{A.}~\bibnamefont{Tyryshkin}},
  \bibinfo{author}{\bibfnamefont{A.}~\bibnamefont{Ardavan}},
  \bibinfo{author}{\bibfnamefont{S.}~\bibnamefont{Benjamin}},
  \bibinfo{author}{\bibfnamefont{K.}~\bibnamefont{Porfyrakis}},
  \bibinfo{author}{\bibfnamefont{S.}~\bibnamefont{Lyon}}, \bibnamefont{and}
  \bibinfo{author}{\bibfnamefont{G.}~\bibnamefont{Briggs}},
  \bibinfo{journal}{Nature Phys.} \textbf{\bibinfo{volume}{2}},
  \bibinfo{pages}{40} (\bibinfo{year}{2006}).

\bibitem[{\citenamefont{Morton et~al.}(2008)\citenamefont{Morton, Tyryshkin,
  Brown, Shankar, Lovett, Ardavan, Schenkel, Haller, Ager, and
  Lyon}}]{Morton:2008}
\bibinfo{author}{\bibfnamefont{J.}~\bibnamefont{Morton}},
  \bibinfo{author}{\bibfnamefont{A.}~\bibnamefont{Tyryshkin}},
  \bibinfo{author}{\bibfnamefont{R.}~\bibnamefont{Brown}},
  \bibinfo{author}{\bibfnamefont{S.}~\bibnamefont{Shankar}},
  \bibinfo{author}{\bibfnamefont{B.}~\bibnamefont{Lovett}},
  \bibinfo{author}{\bibfnamefont{A.}~\bibnamefont{Ardavan}},
  \bibinfo{author}{\bibfnamefont{T.}~\bibnamefont{Schenkel}},
  \bibinfo{author}{\bibfnamefont{E.}~\bibnamefont{Haller}},
  \bibinfo{author}{\bibfnamefont{J.}~\bibnamefont{Ager}}, \bibnamefont{and}
  \bibinfo{author}{\bibfnamefont{S.~A.} \bibnamefont{Lyon}},
  \bibinfo{journal}{Nature} \textbf{\bibinfo{volume}{455}},
  \bibinfo{pages}{1085} (\bibinfo{year}{2008}).

\bibitem[{\citenamefont{Damodarakurup et~al.}()\citenamefont{Damodarakurup,
  Lucamarini, Giuseppe, Vitali, and Tombesi}}]{Vitali2008}
\bibinfo{author}{\bibfnamefont{S.}~\bibnamefont{Damodarakurup}},
  \bibinfo{author}{\bibfnamefont{M.}~\bibnamefont{Lucamarini}},
  \bibinfo{author}{\bibfnamefont{G.~D.} \bibnamefont{Giuseppe}},
  \bibinfo{author}{\bibfnamefont{D.}~\bibnamefont{Vitali}}, \bibnamefont{and}
  \bibinfo{author}{\bibfnamefont{P.}~\bibnamefont{Tombesi}},
\bibinfo{note}{Phys. Rev. Lett. {\bf 103}, 040502 (2009)}.

\bibitem[{Bie()}]{BiercukDD}
\bibinfo{note}{M. J. Biercuk, H. Uys, A. P. VanDevender, N. Shiga, W. M. Itano,
  and J. J. Bollinger, eprint arXiv:0812.5095; eprint arXiv:0902.2957.}

\bibitem[{\citenamefont{Khodjasteh and
  Viola}(2009{\natexlab{a}})}]{khodjasteh-2008}
\bibinfo{author}{\bibfnamefont{K.}~\bibnamefont{Khodjasteh}} \bibnamefont{and}
  \bibinfo{author}{\bibfnamefont{L.}~\bibnamefont{Viola}},
  \bibinfo{journal}{Phys. Rev. Lett.} \textbf{\bibinfo{volume}{102}},
  \bibinfo{pages}{080501} (\bibinfo{year}{2009}{\natexlab{a}}).

\bibitem[{\citenamefont{Levitt}(1986)}]{Levitt1986}
\bibinfo{author}{\bibfnamefont{M.~H.} \bibnamefont{Levitt}},
  \bibinfo{journal}{Progr. NMR Spectrosc.} \textbf{\bibinfo{volume}{18}},
  \bibinfo{pages}{61} (\bibinfo{year}{1986}).

\bibitem[{\citenamefont{Alway and Jones}(2007)}]{Jones-Composite}
\bibinfo{author}{\bibfnamefont{W.~G.} \bibnamefont{Alway}} \bibnamefont{and}
  \bibinfo{author}{\bibfnamefont{J.~A.} \bibnamefont{Jones}},
  \bibinfo{journal}{J. Magn. Res.} \textbf{\bibinfo{volume}{189}},
  \bibinfo{pages}{114} (\bibinfo{year}{2007}).

\bibitem[{\citenamefont{Brown et~al.}(2004)\citenamefont{Brown, Harrow, and
  Chuang}}]{Brown2004}
\bibinfo{author}{\bibfnamefont{K.~R.} \bibnamefont{Brown}},
  \bibinfo{author}{\bibfnamefont{A.~W.} \bibnamefont{Harrow}},
  \bibnamefont{and} \bibinfo{author}{\bibfnamefont{I.~L.}
  \bibnamefont{Chuang}}, \bibinfo{journal}{Phys. Rev. A}
  \textbf{\bibinfo{volume}{70}}, \bibinfo{pages}{052318}
  (\bibinfo{year}{2004}).

\bibitem[{\citenamefont{Khaneja et~al.}(2005)\citenamefont{Khaneja, Reiss,
  Kehlet, Schulte-Herbr\"uggen, and Glaser}}]{KhanejaUR}
\bibinfo{author}{\bibfnamefont{N.}~\bibnamefont{Khaneja}},
  \bibinfo{author}{\bibfnamefont{T.}~\bibnamefont{Reiss}},
  \bibinfo{author}{\bibfnamefont{C.}~\bibnamefont{Kehlet}},
  \bibinfo{author}{\bibfnamefont{T.}~\bibnamefont{Schulte-Herbr\"uggen}},
  \bibnamefont{and} \bibinfo{author}{\bibfnamefont{S.~J.}
  \bibnamefont{Glaser}}, \bibinfo{journal}{J. Magn. Res.}
  \textbf{\bibinfo{volume}{172}}, \bibinfo{pages}{296} (\bibinfo{year}{2005}).

\bibitem[{\citenamefont{Hill}(2007)}]{Hill2007}
\bibinfo{author}{\bibfnamefont{C.~D.} \bibnamefont{Hill}},
  \bibinfo{journal}{Phys. Rev. Lett.} \textbf{\bibinfo{volume}{98}},
  \bibinfo{pages}{180501} (\bibinfo{year}{2007}).

\bibitem[{\citenamefont{Fortunato et~al.}(2002)\citenamefont{Fortunato, Pravia,
  Boulant, Teklemariam, Havel, and Cory}}]{Fortunato-Control}
\bibinfo{author}{\bibfnamefont{E.~M.} \bibnamefont{Fortunato}},
  \bibinfo{author}{\bibfnamefont{M.~A.} \bibnamefont{Pravia}},
  \bibinfo{author}{\bibfnamefont{N.}~\bibnamefont{Boulant}},
  \bibinfo{author}{\bibfnamefont{G.}~\bibnamefont{Teklemariam}},
  \bibinfo{author}{\bibfnamefont{T.~F.} \bibnamefont{Havel}}, \bibnamefont{and}
  \bibinfo{author}{\bibfnamefont{D.~G.} \bibnamefont{Cory}},
  \bibinfo{journal}{J. Chem. Phys.} \textbf{\bibinfo{volume}{116}},
  \bibinfo{pages}{7599} (\bibinfo{year}{2002}).

\bibitem[{\citenamefont{Boulant et~al.}(2003)\citenamefont{Boulant, Edmonds,
  Yang, Pravia, and Cory}}]{Boulant03}
\bibinfo{author}{\bibfnamefont{N.}~\bibnamefont{Boulant}},
  \bibinfo{author}{\bibfnamefont{K.}~\bibnamefont{Edmonds}},
  \bibinfo{author}{\bibfnamefont{J.}~\bibnamefont{Yang}},
  \bibinfo{author}{\bibfnamefont{M.~A.} \bibnamefont{Pravia}},
  \bibnamefont{and} \bibinfo{author}{\bibfnamefont{D.~G.} \bibnamefont{Cory}},
  \bibinfo{journal}{Phys. Rev. A} \textbf{\bibinfo{volume}{68}},
  \bibinfo{pages}{032305} (\bibinfo{year}{2003}).

\bibitem[{\citenamefont{Luy et~al.}(2005)\citenamefont{Luy, Kozbar, Skinner,
  Khaneja, and Glaser}}]{KhanejaGRAPE}
\bibinfo{author}{\bibfnamefont{B.}~\bibnamefont{Luy}},
  \bibinfo{author}{\bibfnamefont{K.}~\bibnamefont{Kozbar}},
  \bibinfo{author}{\bibfnamefont{T.~E.} \bibnamefont{Skinner}},
  \bibinfo{author}{\bibfnamefont{N.}~\bibnamefont{Khaneja}}, \bibnamefont{and}
  \bibinfo{author}{\bibfnamefont{S.~J.} \bibnamefont{Glaser}},
  \bibinfo{journal}{J. Magn. Res.} \textbf{\bibinfo{volume}{179}},
  \bibinfo{pages}{176} (\bibinfo{year}{2005}).

\bibitem[{\citenamefont{Pasini et~al.}(2008)\citenamefont{Pasini, Fischer,
  Karbach, and Uhrig}}]{Pasini07}
\bibinfo{author}{\bibfnamefont{S.}~\bibnamefont{Pasini}},
  \bibinfo{author}{\bibfnamefont{T.}~\bibnamefont{Fischer}},
  \bibinfo{author}{\bibfnamefont{P.}~\bibnamefont{Karbach}}, \bibnamefont{and}
  \bibinfo{author}{\bibfnamefont{G.~S.} \bibnamefont{Uhrig}},
  \bibinfo{journal}{Phys. Rev. A} \textbf{\bibinfo{volume}{77}},
  \bibinfo{pages}{032315} (\bibinfo{year}{2008}).

\bibitem[{\citenamefont{Pryadko and Quiroz}(2008)}]{PryadkoQuiroz}
\bibinfo{author}{\bibfnamefont{L.~P.} \bibnamefont{Pryadko}} \bibnamefont{and}
  \bibinfo{author}{\bibfnamefont{G.}~\bibnamefont{Quiroz}},
  \bibinfo{journal}{Phys. Rev. A} \textbf{\bibinfo{volume}{77}},
  \bibinfo{pages}{012330} (\bibinfo{year}{2008}).

\bibitem[{\citenamefont{Sengupta and Pryadko}(2005)}]{Sengupta05}
\bibinfo{author}{\bibfnamefont{P.}~\bibnamefont{Sengupta}} \bibnamefont{and}
  \bibinfo{author}{\bibfnamefont{L.~P.} \bibnamefont{Pryadko}},
  \bibinfo{journal}{Phys. Rev. Lett.} \textbf{\bibinfo{volume}{95}},
  \bibinfo{pages}{037202} (\bibinfo{year}{2005}).

\bibitem[{\citenamefont{Viola and Knill}(2003)}]{Viola2003Euler}
\bibinfo{author}{\bibfnamefont{L.}~\bibnamefont{Viola}} \bibnamefont{and}
  \bibinfo{author}{\bibfnamefont{E.}~\bibnamefont{Knill}},
  \bibinfo{journal}{Phys. Rev. Lett.} \textbf{\bibinfo{volume}{90}},
  \bibinfo{pages}{037901} (\bibinfo{year}{2003}).

\bibitem[{\citenamefont{Terhal and Burkard}(2005)}]{Terhal-FT}
\bibinfo{author}{\bibfnamefont{B.~M.}~\bibnamefont{Terhal}} \bibnamefont{and}
  \bibinfo{author}{\bibfnamefont{G.}~\bibnamefont{Burkard}},
  \bibinfo{journal}{Phys. Rev. A} \textbf{\bibinfo{volume}{71}},
  \bibinfo{pages}{012336} (\bibinfo{year}{2005}).

\bibitem[{\citenamefont{Khodjasteh and
  Viola}(2009{\natexlab{b}})}]{khodjasteh-2009}
\bibinfo{author}{\bibfnamefont{K.}~\bibnamefont{Khodjasteh}},
   D. A. Lidar, \bibnamefont{and}
  \bibinfo{author}{\bibfnamefont{L.}~\bibnamefont{Viola}}, eprint arXiv:0908.1526.

\bibitem[{\citenamefont{Zurek}(2003)}]{RevModPhys.75.715}
\bibinfo{author}{\bibfnamefont{W.~H.} \bibnamefont{Zurek}},
  \bibinfo{journal}{Rev. Mod. Phys.} \textbf{\bibinfo{volume}{75}},
  \bibinfo{pages}{715} (\bibinfo{year}{2003}).

\bibitem[{\citenamefont{Breuer and Petruccione}(2002)}]{Breuer}
\bibinfo{author}{\bibfnamefont{H.-P.} \bibnamefont{Breuer}} \bibnamefont{and}
  \bibinfo{author}{\bibfnamefont{F.}~\bibnamefont{Petruccione}},
  \emph{\bibinfo{title}{The Theory of Open Quantum Systems}}
  (\bibinfo{publisher}{Oxford University Press}, \bibinfo{address}{Oxford},
  \bibinfo{year}{2002}).

\bibitem[{\citenamefont{Knill et~al.}(2000)\citenamefont{Knill, Laflamme, and
  Viola}}]{Knill-NS}
\bibinfo{author}{\bibfnamefont{E.}~\bibnamefont{Knill}},
  \bibinfo{author}{\bibfnamefont{R.}~\bibnamefont{Laflamme}}, \bibnamefont{and}
  \bibinfo{author}{\bibfnamefont{L.}~\bibnamefont{Viola}},
  \bibinfo{journal}{Phys. Rev. Lett.} \textbf{\bibinfo{volume}{84}},
  \bibinfo{pages}{2525} (\bibinfo{year}{2000}).

\bibitem[{\citenamefont{Bhatia}(1997)}]{bhatiaBook}
\bibinfo{author}{\bibfnamefont{R.}~\bibnamefont{Bhatia}},
  \emph{\bibinfo{title}{Matrix Analysis}}, no. \bibinfo{number}{169} in
  \bibinfo{series}{Graduate Texts in Mathematics}
  (\bibinfo{publisher}{Springer-Verlag, New York}, \bibinfo{year}{1997}).

\bibitem[{\citenamefont{Lidar et~al.}(2008)\citenamefont{Lidar, Zanardi, and
  Khodjasteh}}]{Lidar-Bounds}
\bibinfo{author}{\bibfnamefont{D.~A.} \bibnamefont{Lidar}},
  \bibinfo{author}{\bibfnamefont{P.}~\bibnamefont{Zanardi}}, \bibnamefont{and}
  \bibinfo{author}{\bibfnamefont{K.}~\bibnamefont{Khodjasteh}},
  \bibinfo{journal}{Phys. Rev. A} \textbf{\bibinfo{volume}{78}},
  \bibinfo{pages}{012308} (\bibinfo{year}{2008}).

\bibitem[{\citenamefont{Khodjasteh and Lidar}(2008)}]{Khodjasteh-Hybrid}
\bibinfo{author}{\bibfnamefont{K.}~\bibnamefont{Khodjasteh}} \bibnamefont{and}
  \bibinfo{author}{\bibfnamefont{D.~A.} \bibnamefont{Lidar}},
  \bibinfo{journal}{Phys. Rev. A} \textbf{\bibinfo{volume}{78}},
  \bibinfo{pages}{012355} (\bibinfo{year}{2008}).

\bibitem[{\citenamefont{D'Alessandro}(2007)}]{D'Alessandro}
\bibinfo{author}{\bibfnamefont{D.}~\bibnamefont{D'Alessandro}},
  \emph{\bibinfo{title}{Introduction to Quantum Control and Dynamics}}
  (\bibinfo{publisher}{CRC Press}, \bibinfo{address}{Boca Raton},
  \bibinfo{year}{2007}).

\bibitem[{\citenamefont{Khodjasteh and Lidar}(2005)}]{Khodjasteh2004}
\bibinfo{author}{\bibfnamefont{K.}~\bibnamefont{Khodjasteh}} \bibnamefont{and}
  \bibinfo{author}{\bibfnamefont{D.~A.} \bibnamefont{Lidar}},
  \bibinfo{journal}{Phys. Rev. Lett.} \textbf{\bibinfo{volume}{95}},
  \bibinfo{pages}{180501} (\bibinfo{year}{2005}).

\bibitem[{\citenamefont{Uhrig}(2007)}]{Uhrig2007}
\bibinfo{author}{\bibfnamefont{G.~S.} \bibnamefont{Uhrig}},
  \bibinfo{journal}{Phys. Rev. Lett.} \textbf{\bibinfo{volume}{98}},
  \bibinfo{pages}{100504} (\bibinfo{year}{2007}).

\bibitem[{\citenamefont{Uys et~al.}()\citenamefont{Uys, Biercuk, and
  Bollinger}}]{Uys09}
\bibinfo{author}{\bibfnamefont{H.}~\bibnamefont{Uys}},
  \bibinfo{author}{\bibfnamefont{M.~J.} \bibnamefont{Biercuk}},
  \bibnamefont{and} \bibinfo{author}{\bibfnamefont{J.~J.}
  \bibnamefont{Bollinger}}, \bibinfo{note}{arXiv:0904.0036}.

\bibitem[{\citenamefont{Khodjasteh and Lidar}(2007)}]{Khodjasteh2005}
\bibinfo{author}{\bibfnamefont{K.}~\bibnamefont{Khodjasteh}} \bibnamefont{and}
  \bibinfo{author}{\bibfnamefont{D.~A.} \bibnamefont{Lidar}},
  \bibinfo{journal}{Phys. Rev. A} \textbf{\bibinfo{volume}{75}},
  \bibinfo{pages}{062310} (\bibinfo{year}{2007}).

\bibitem[{\citenamefont{Santos and Viola}(2008)}]{Santos2008}
\bibinfo{author}{\bibfnamefont{L.~F.} \bibnamefont{Santos}} \bibnamefont{and}
  \bibinfo{author}{\bibfnamefont{L.}~\bibnamefont{Viola}},
  \bibinfo{journal}{New J. Phys.} \textbf{\bibinfo{volume}{10}},
  \bibinfo{pages}{083009} (\bibinfo{year}{2008}).

\bibitem[{\citenamefont{R\"otteler and Wocjan}(2006)}]{Wocjan2006}
\bibinfo{author}{\bibfnamefont{M.}~\bibnamefont{R\"otteler}} \bibnamefont{and}
  \bibinfo{author}{\bibfnamefont{P.}~\bibnamefont{Wocjan}},
  \bibinfo{journal}{IEEE Trans. Inf. Theory} \textbf{\bibinfo{volume}{52}},
  \bibinfo{pages}{4171} (\bibinfo{year}{2006}).

\bibitem[{\citenamefont{Viola et~al.}(2000)\citenamefont{Viola, Knill, and
  Lloyd}}]{Viola2000Dygen}
\bibinfo{author}{\bibfnamefont{L.}~\bibnamefont{Viola}},
  \bibinfo{author}{\bibfnamefont{E.}~\bibnamefont{Knill}}, \bibnamefont{and}
  \bibinfo{author}{\bibfnamefont{S.}~\bibnamefont{Lloyd}},
  \bibinfo{journal}{Phys. Rev. Lett.} \textbf{\bibinfo{volume}{85}},
  \bibinfo{pages}{3520} (\bibinfo{year}{2000}).

\bibitem[{\citenamefont{Viola}(2002)}]{Viola2002}
\bibinfo{author}{\bibfnamefont{L.}~\bibnamefont{Viola}},
  \bibinfo{journal}{Phys. Rev. A} \textbf{\bibinfo{volume}{66}},
  \bibinfo{pages}{012307} (\bibinfo{year}{2002}).

\bibitem[{\citenamefont{Lidar}(2008)}]{lidar:160506}
\bibinfo{author}{\bibfnamefont{D.~A.} \bibnamefont{Lidar}},
  \bibinfo{journal}{Phys. Rev. Lett.} \textbf{\bibinfo{volume}{100}},
  \bibinfo{pages}{160506} (\bibinfo{year}{2008}).

\bibitem[{Not({\natexlab{a}})}]{Note4}
\bibinfo{note}{Within the radius of convergence, the Magnus expansion is
  absolutely convergent. This fact can be used to cap the sum of the higher
  order terms, see {\em e.g.} \cite{Khodjasteh2005}.}

\bibitem[{\citenamefont{Aliferis and Preskill}(2008)}]{aliferis:08}
\bibinfo{author}{\bibfnamefont{P.}~\bibnamefont{Aliferis}} \bibnamefont{and}
  \bibinfo{author}{\bibfnamefont{J.}~\bibnamefont{Preskill}},
  \bibinfo{journal}{Phys. Rev. A} \textbf{\bibinfo{volume}{78}},
  \bibinfo{pages}{052331} (\bibinfo{year}{2008}).

\bibitem[{Not({\natexlab{b}})}]{Note5}
\bibinfo{note}{The so-called Uhrig DD is an example where the algebraic
  simplicity of the dephasing model is exploited to obtain an {\em efficient}
  DQEC protocol \cite{Uhrig2007}.}

\bibitem[{\citenamefont{Viola}(2004)}]{Viola2004}
\bibinfo{author}{\bibfnamefont{L.}~\bibnamefont{Viola}}, \bibinfo{journal}{J.
  Mod. Opt.} \textbf{\bibinfo{volume}{51}}, \bibinfo{pages}{2357}
  (\bibinfo{year}{2004}).

\bibitem[{Nir()}]{Nir}
\bibinfo{note}{Interestingly, a close-in-spirit implementation has been
  recently reported for dynamical suppression of collisional decoherence in
  cold trapped atoms, see Y. Sagi, I. Almog and N. Davidson, eprint
  arXiv:0905.0286.}

\bibitem[{\citenamefont{Viola et~al.}(1999{\natexlab{b}})\citenamefont{Viola,
  Lloyd, and Knill}}]{Viola1999Control}
\bibinfo{author}{\bibfnamefont{L.}~\bibnamefont{Viola}},
  \bibinfo{author}{\bibfnamefont{S.}~\bibnamefont{Lloyd}}, \bibnamefont{and}
  \bibinfo{author}{\bibfnamefont{E.}~\bibnamefont{Knill}},
  \bibinfo{journal}{Phys. Rev. Lett.} \textbf{\bibinfo{volume}{83}},
  \bibinfo{pages}{4888} (\bibinfo{year}{1999}{\natexlab{b}}).

\bibitem[{\citenamefont{Loss and DiVincenzo}(1998)}]{Loss}
\bibinfo{author}{\bibfnamefont{D.}~\bibnamefont{Loss}} \bibnamefont{and}
  \bibinfo{author}{\bibfnamefont{D.~P.} \bibnamefont{DiVincenzo}},
  \bibinfo{journal}{Phys. Rev. A} \textbf{\bibinfo{volume}{57}},
  \bibinfo{pages}{120} (\bibinfo{year}{1998}).

\bibitem[{\citenamefont{Hanson et~al.}(2007)\citenamefont{Hanson, Kouwenhoven,
  Petta, Tarucha, and Vandersypen}}]{HansonRev}
\bibinfo{author}{\bibfnamefont{R.}~\bibnamefont{Hanson}},
  \bibinfo{author}{\bibfnamefont{L.~P.} \bibnamefont{Kouwenhoven}},
  \bibinfo{author}{\bibfnamefont{J.~R.} \bibnamefont{Petta}},
  \bibinfo{author}{\bibfnamefont{S.}~\bibnamefont{Tarucha}}, \bibnamefont{and}
  \bibinfo{author}{\bibfnamefont{L.~M.~K.} \bibnamefont{Vandersypen}},
  \bibinfo{journal}{Rev. Mod. Phys.} \textbf{\bibinfo{volume}{79}},
  \bibinfo{pages}{1217} (\bibinfo{year}{2007}).

\bibitem[{\citenamefont{Zhang et~al.}(2007{\natexlab{a}})\citenamefont{Zhang,
  Konstantinidis, Al-Hassanieh, and Dobrovitski}}]{WenReview}
\bibinfo{author}{\bibfnamefont{W.}~\bibnamefont{Zhang}},
  \bibinfo{author}{\bibfnamefont{N.~P.} \bibnamefont{Konstantinidis}},
  \bibinfo{author}{\bibfnamefont{K.~A.} \bibnamefont{Al-Hassanieh}},
  \bibnamefont{and} \bibinfo{author}{\bibfnamefont{V.~V.}
  \bibnamefont{Dobrovitski}}, \bibinfo{journal}{J. Phys.: Cond. Matter}
  \textbf{\bibinfo{volume}{19}}, \bibinfo{pages}{083202}
  (\bibinfo{year}{2007}{\natexlab{a}}).

\bibitem[{Not({\natexlab{c}})}]{Note6}
\bibinfo{note}{Identical DCG constructions would be valid if the Heisenberg
  interaction is replaced by any two-qubit Hamiltonian commuting with
  $S_{\alpha}^{(\text{all})}$ ({\em e.g.}, the Ising interaction).}

\bibitem[{Not({\natexlab{d}})}]{Note7}
\bibinfo{note}{In principle, the unit step function could be replaced by any
  function $u(x)$ such that $u(x)=0\text{ if } x \notin[0,1]$ and
  $\int_0^{1}u(x)dx=1$. Even different gates need not be based on the same
  profile $u(x)$. The essential requirement, however, is that the control
  profiles are flexible enough to allow the constructions of $Q'$ and $Q_{1/2}$
  in Sec. \ref{sub:diffequal}.}

\bibitem[{\citenamefont{Zhang et~al.}(2007{\natexlab{b}})\citenamefont{Zhang,
  Dobrovitski, Santos, Viola, and Harmon}}]{Wen01}
\bibinfo{author}{\bibfnamefont{W.}~\bibnamefont{Zhang}},
  \bibinfo{author}{\bibfnamefont{V.~V.} \bibnamefont{Dobrovitski}},
  \bibinfo{author}{\bibfnamefont{L.~F.} \bibnamefont{Santos}},
  \bibinfo{author}{\bibfnamefont{L.}~\bibnamefont{Viola}}, \bibnamefont{and}
  \bibinfo{author}{\bibfnamefont{B.~N.} \bibnamefont{Harmon}},
  \bibinfo{journal}{Phys. Rev. B} \textbf{\bibinfo{volume}{75}},
  \bibinfo{pages}{201302(R)} (\bibinfo{year}{2007}{\natexlab{b}}).

\bibitem[{\citenamefont{Zhang et~al.}(2007{\natexlab{c}})\citenamefont{Zhang,
  Dobrovitski, Santos, Viola, and Harmon}}]{Wen02}
\bibinfo{author}{\bibfnamefont{W.}~\bibnamefont{Zhang}},
  \bibinfo{author}{\bibfnamefont{V.~V.} \bibnamefont{Dobrovitski}},
  \bibinfo{author}{\bibfnamefont{L.~F.} \bibnamefont{Santos}},
  \bibinfo{author}{\bibfnamefont{L.}~\bibnamefont{Viola}}, \bibnamefont{and}
  \bibinfo{author}{\bibfnamefont{B.~N.} \bibnamefont{Harmon}},
  \bibinfo{journal}{J. Mod. Opt.} \textbf{\bibinfo{volume}{54}},
  \bibinfo{pages}{2629} (\bibinfo{year}{2007}{\natexlab{c}}).

\bibitem[{\citenamefont{Zhang et~al.}(2008)\citenamefont{Zhang, Konstantinidis,
  Dobrovitski, Harmon, Santos, and Viola}}]{Wen03}
\bibinfo{author}{\bibfnamefont{W.}~\bibnamefont{Zhang}},
  \bibinfo{author}{\bibfnamefont{N.~P.} \bibnamefont{Konstantinidis}},
  \bibinfo{author}{\bibfnamefont{V.~V.} \bibnamefont{Dobrovitski}},
  \bibinfo{author}{\bibfnamefont{B.~N.} \bibnamefont{Harmon}},
  \bibinfo{author}{\bibfnamefont{L.~F.} \bibnamefont{Santos}},
  \bibnamefont{and} \bibinfo{author}{\bibfnamefont{L.}~\bibnamefont{Viola}},
  \bibinfo{journal}{Phys. Rev. B} \textbf{\bibinfo{volume}{77}},
  \bibinfo{pages}{125336} (\bibinfo{year}{2008}).

\bibitem[{Rea()}]{Realizations}
\bibinfo{note}{Realization-to-realization fluctuations in the improvement ratio
  and corresponding fidelities have been found to be more or less pronounced
  depending on parameter regime. In particular, the variations are largest for
  $r \lesssim 1$ and large $A$ (yielding {\em e.g.} up to a $20\%$ relative
  variance over $4$ different samples), whereas the effect of $\Gamma$ seems
  far less important. These fluctuations do not in any case qualitatively alter
  the main observed features.}

\bibitem[{\citenamefont{Cummins and Jones}(2000)}]{JonesComp}
\bibinfo{author}{\bibfnamefont{H.~K.} \bibnamefont{Cummins}} \bibnamefont{and}
  \bibinfo{author}{\bibfnamefont{J.~A.} \bibnamefont{Jones}},
  \bibinfo{journal}{New J. Phys.} \textbf{\bibinfo{volume}{2}},
  \bibinfo{pages}{6.1} (\bibinfo{year}{2000}).

\bibitem[{\citenamefont{Hsieh and Rabitz}(2008)}]{RabitzLand}
\bibinfo{author}{\bibfnamefont{M.}~\bibnamefont{Hsieh}} \bibnamefont{and}
  \bibinfo{author}{\bibfnamefont{H.}~\bibnamefont{Rabitz}},
  \bibinfo{journal}{Phys. Rev. A} \textbf{\bibinfo{volume}{77}},
  \bibinfo{pages}{042306} (\bibinfo{year}{2008}).

\bibitem[{\citenamefont{Levy et~al.}()\citenamefont{Levy, Ganti, Phillips,
  Hamlet, Landahl, Gurrieri, Carr, and Carroll}}]{Levy09}
\bibinfo{author}{\bibfnamefont{J.~E.} \bibnamefont{Levy}},
  \bibinfo{author}{\bibfnamefont{A.}~\bibnamefont{Ganti}},
  \bibinfo{author}{\bibfnamefont{C.~A.} \bibnamefont{Phillips}},
  \bibinfo{author}{\bibfnamefont{B.~R.} \bibnamefont{Hamlet}},
  \bibinfo{author}{\bibfnamefont{A.~J.} \bibnamefont{Landahl}},
  \bibinfo{author}{\bibfnamefont{T.~M.} \bibnamefont{Gurrieri}},
  \bibinfo{author}{\bibfnamefont{R.~D.} \bibnamefont{Carr}}, \bibnamefont{and}
  \bibinfo{author}{\bibfnamefont{M.~S.} \bibnamefont{Carroll}},
  \bibinfo{note}{eprint arXiv:0904.0003}.

\end{thebibliography}
\end{document}